\documentclass[12pt,a4paper]{article} 
\usepackage{style}
\usepackage{amsmath}
\usepackage{amssymb}
\usepackage{amsthm}
\usepackage{psfrag}
\usepackage{graphicx}
\usepackage{hyperref}
\usepackage{feynmp}
\usepackage{color}
\usepackage{bm}

\renewcommand{\L}{\mathcal{L}}

\renewcommand{\l}{\lambda}
\renewcommand{\d}{\partial}
\renewcommand{\a}{\alpha}

\newcommand{\be}{\begin{equation}}
\newcommand{\ee}{\end{equation}}
\newcommand{\beqa}{\begin{eqnarray}}
\newcommand{\eeqa}{\end{eqnarray}}
\newcommand{\bsm}{\begin{smallmatrix}}
\newcommand{\esm}{\end{smallmatrix}}

\renewcommand\r{\rho}

\renewcommand\a{\alpha}

\renewcommand\l{\lambda}

\newcommand\x{{\bf x}}
\renewcommand\k{{\bf k}}
\newcommand\q{{\bf q}}

\newcommand{\e}{\eta}
\newcommand{\hMpc}{h\text{Mpc}^{-1}}

\def\e{{\rm e}}
\def\d{\partial}
\newcommand{\bseq}{\begin{subequations}}
\newcommand{\eseq}{\end{subequations}}

\renewcommand{\ln}{\mathop{\rm ln}\nolimits}

\renewcommand{\H}{\mathcal H}

\renewcommand{\L}{\Lambda}

\renewcommand{\k}{{\bf k}}

\newcommand{\z}{{\bf z}}
\newcommand{\kmax}{k_{\rm max}}

\renewcommand{\d}{\partial}

\def\l{\left(}
\def\r{\right)}

\newcommand{\astfootnote}[1]{%
\let\oldthefootnote=\thefootnote%
\setcounter{footnote}{0}%
\renewcommand{\thefootnote}{\fnsymbol{footnote}}%
\footnote{#1}%
\let\thefootnote=\oldthefootnote%
}

\relax

\title{Precision analysis of the redshift-space galaxy bispectrum}

\author[a]{Mikhail M. Ivanov,\footnote{\texttt{ivanov@ias.edu}}\astfootnote{Einstein Fellow}}
\author[a,b]{Oliver H.\,E. Philcox,}
\author[c,d]{Takahiro Nishimichi,} 
\author[e]{Marko Simonovi\'c,}
\author[d]{Masahiro Takada,}
\author[a]{and Matias Zaldarriaga}
\affiliation[a]{School of Natural Sciences, Institute for Advanced Study,\\1 Einstein Drive, Princeton, NJ 08540, USA}
\affiliation[b]{Department of Astrophysical Sciences, Princeton University,\\ Princeton, NJ 08540, USA}%
\affiliation[c]{Center for Gravitational Physics, \\ Yukawa Institute for Theoretical Physics, Kyoto University, Kyoto 606-8502, Japan}
\affiliation[d]{Kavli Institute for the Physics and Mathematics of the Universe (WPI), UTIAS \\The University of Tokyo, Kashiwa, Chiba 277-8583, Japan}
\affiliation[e]{Theoretical Physics Department, CERN,\\1 Esplanade des Particules, Geneva 23, CH-1211, Switzerland}

\abstract{
We study the information content of 
the angle-averaged (monopole) redshift space galaxy bispectrum. 
The main novelty of our approach 
is the use of a systematic
tree-level perturbation theory model that includes
galaxy bias, IR resummation, 
and also accounts for nonlinear redshift space distortions, binning,
and projection effects.
We analyze data
from 
the PT challenge simulations, 
whose cumulative volume of 566~$h^{-3}$Gpc$^3$ allows for a 
precise
comparison to
theoretical
predictions.
Fitting the power spectrum and bispectrum
of our simulated data,
and varying all necessary cosmological 
and nuisance parameters in a consistent Markov chain Monte Carlo analysis,
we find that 
our tree-level bispectrum
model
is valid 
up to $\kmax=0.08~\hMpc$ (at $z=0.61$).
We also find that inclusion of the bispectrum monopole
improves constraints on cosmological parameters by $(5-15)\%$ relative to the power spectrum.
The improvement is more significant
for the quadratic bias parameters
of our simulated galaxies, 
which we also show to deviate from 
biases of the host dark matter halos at the $\sim 3\sigma$ level.
Finally, we adjust the covariance and scale cuts 
to match the volume of the BOSS survey,
and estimate that within the minimal 
$\Lambda$CDM model
the bispectrum data can tighten  
the constraint on 
the mass fluctuation amplitude $\sigma_8$ by roughly $10\%$.
}

\begin{document}

\begin{flushright}
YITP-21-120,
% 	% INR-TH-2020-xxx 
	CERN-TH-2021-155
\end{flushright}

\maketitle

\section{Introduction}

The three-point function, or its Fourier transform the bispectrum~\cite{1980lssu.book.....P}, is the simplest statistic beyond the power spectrum that captures information about the large-scale spatial distribution of galaxies. The shape dependence of the bispectrum is sensitive to cosmological initial conditions, gravitational instability, and galaxy formation physics. For this reason the bispectrum is an important 
observational probe which can improve our understanding of both galaxy formation and fundamental cosmology~\cite{Scoccimarro:2000sn,Sefusatti:2006pa}. It has been argued that it may help sharpen the limits on conventional cosmological parameters~\cite{
Sefusatti:2006pa,
Baldauf:2016sjb,Yankelevich:2018uaz,Chudaykin:2019ock}, neutrino 
masses~\cite{Chudaykin:2019ock,Hahn:2019zob,Hahn:2020lou},
and primordial non-Gaussianity~\cite{Welling:2016dng,Baldauf:2016sjb,MoradinezhadDizgah:2020whw}. While these results are encouraging, they are often based on idealized Fisher forecasts 
and overoptimistic assumptions about the
validity of theoretical models needed to describe the data. 
Therefore, it is still
not clear 
whether the inclusion of the bispectrum
will be worth the effort, 
whether it will really make a difference 
in a realistic analysis 
where all relevant cosmological and nuisance parameters are varied.

A quantitative answer to this question cannot be given without performing a consistent data analysis. While the three-point functions and bispectra of the galaxy density field 
have been 
measured both in simulations 
and in a number 
of past and current datasets (e.g.~Zwicky and
Lick catalogs~\cite{1975ApJ...196....1P,1977ApJ...217..385G}, IRAS~\cite{Scoccimarro:2000sp,Feldman:2000vk}, 
WiggleZ~\cite{Marin:2013bbb},  Baryon Oscillation
Spectroscopic Survey (BOSS)~\cite{Gil-Marin:2014sta,Gil-Marin:2014baa,Gil-Marin:2016wya,Slepian:2015hca}), the proper cosmological analyses of the bispectrum are still lacking. This is clearly in sharp contrast with the galaxy power spectrum analyses, which have been routinely used as an important source of information on cosmological parameters.
There are multiple factors that make the bispectrum analysis much more challenging.

From the computational side, the main challenge is a large number 
of data points, which correspond to triangle configurations formed by 
three wavevectors $\k_1,\k_2,\k_3$. Typical bispectra datasets consist 
of hundreds of triangles, which makes it hard 
to estimate the bispectrum from catalogs,
compute the covariance matrix, and perform likelihood analysis.
This stimulated the development of 
fast estimators~\cite{Scoccimarro:2015bla,Philcox:2019hdi,Philcox:2019xzt,Philcox:2020xod,Philcox:2021bwo},
various compression techniques~\cite{Scoccimarro:2000sn,Gaztanaga:2005ad,Gualdi:2018pyw,Philcox:2020zyp} 
and efficient mock catalog pipelines~\cite{Kitaura:2015uqa}.

From the theory side, the main challenge is modeling non-linear effects of matter clustering, galaxy bias, and redshift space distortions.
Recent analyses described these effects 
by means of N-body simulations, which were used to calibrate 
phenomenological bispectrum models~\cite{Kitaura:2014mja,Gil-Marin:2014sta,Gil-Marin:2014baa,Gil-Marin:2016wya,Takahashi_2020}.
This simulation-based approach naturally extends to 
`emulation', 
in which the data is fitted 
directly to the simulation output~\cite{Heitmann:2009cu,2019ApJ...884...29N,2020PhRvD.102f3504K,Hahn:2019zob,Hahn:2020lou}. 
Despite significant progress 
in numerical modeling of galaxy clustering over last years, 
it is not yet clear if emulators can meet precision 
requirements of future surveys, see e.g.~\cite{Schneider:2015yka}.
The main issue is persistent uncertainty in galaxy formation physics, 
which has to be marginalized over in order to obtain robust 
cosmological constraints.
This motivates the development of more 
conservative 
perturbative techniques~\cite{Fry:1983cj,Scoccimarro:1995if,Scoccimarro:1997st,Scoccimarro:1999ed}, which have recently taken non-linear 
large-scale structure 
modeling 
to a new  precision level by virtue of the progress in the effective field 
theory (EFT) of large-scale structure~\cite{Baumann:2010tm,Carrasco:2012cv}.\footnote{In what follows we will not distinguish between perturbation theory and the EFT, as the EFT is the only 
consistent realization of large-scale structure perturbation theory.}

Unlike simulation-based approaches, EFT  
is fundamentally restricted to scales
larger than $2\pi k_{\rm NL}^{-1}\sim 10$ Mpc. 
However, in the regime where it is applicable, 
EFT allows
calculations to arbitrary order, and hence 
it provides a program of 
systematic successive approximations
to the true answer.
Moreover, by construction,
EFT covers all possible galaxy formation scenarios 
by means 
of ``nuisance parameters,'' which fully capture the impact 
of galaxy evolution on large scale clustering.
Thus, this framework is naturally designed for the marginalization 
over galaxy formation physics, 
which boils down to a literal marginalization over nuisance 
parameters. Finally, EFT-based theoretical templates 
for a given cosmological model
can be quickly generated with modifications of Boltzmann codes, e.g.~\cite{Chudaykin:2020aoj,DAmico:2020kxu,Chen:2020zjt},
which allow one to efficiently explore the cosmology-dependence 
of large-scale structure data.

The full utility of the EFT approach has been shown recently in the analysis 
of the galaxy power spectrum data from BOSS~\cite{Alam:2016hwk}.
This has resulted in first-ever measurements 
of \textit{fundamental} cosmological parameters, such as the Hubble 
constant and the amplitude of the primordial scalar fluctuations,
from the full shape of the galaxy power 
spectrum~\cite{Ivanov:2019pdj,DAmico:2019fhj}.
Moreover, the EFT-based full shape analyses have opened up a new opportunity 
to testing beyond-$\Lambda$CDM scenarios in a rigorous and self-consistent
fashion~\cite{Ivanov:2019hqk,Ivanov:2020ril,DAmico:2020kxu,DAmico:2020ods,Chudaykin:2020ghx}. 

An important step in applying the EFT calculations to the real data
was the validation of the EFT-based power spectrum likelihoods 
on high-fidelity simulations~\cite{Ivanov:2019pdj,DAmico:2019fhj,Chudaykin:2020hbf,Chudaykin:2020ghx}. 
In particular, the EFT-based pipelines have passed a blind test 
on galaxy mock catalogs called ``PT challenge''~\cite{Nishimichi:2020tvu}.\footnote{The aim of this challenge is to test various methods of cosmological parameter inference from large-scale structure data
in a blind way.
The Reader is welcome to participate. The challenge details can be found at~\url{https://www2.yukawa.kyoto-u.ac.jp/~takahiro.nishimichi/data/PTchallenge/}}  
The PT challenge simulation
suite covers a cumulative volume of $566~h^{-3}$Gpc$^3$, which is significantly larger
than the volume of current and planned surveys. This large volume is chosen 
with the purpose of dramatically reducing statistical error 
and thereby identifying
systematic uncertainties 
in theoretical modeling at the unprecedented sub-percent level.

Inspired by the success of the EFT approach in the power spectrum analysis, in this work we extend the study of 
the PT challenge simulation data 
from Ref.~\cite{Nishimichi:2020tvu}
to the galaxy bispectrum.
We analyze this data with the currently available tree-level EFT model.\footnote{Perturbation theory one-loop bispectra 
of matter and halos in real space
% in 
have been studied in Refs.~\cite{Valageas_2011,Angulo:2014tfa,Baldauf:2014qfa,Bertolini:2016bmt,Eggemeier:2018qae,Taruya_2018,Osato_2021}.
While these calculations have not yet been extended to the realistic 
case of galaxy clustering in redshift space,
certain relevant ingredients are already 
available in the literature, e.g.~the redshift-space mapping in the EFT~\cite{Senatore:2014vja,Lewandowski:2015ziq,Perko:2016puo},
the perturbative bias model~\cite{Senatore:2014eva,Angulo:2015eqa,Assassi:2014fva,Mirbabayi:2014zca,Desjacques:2016bnm,Desjacques:2018pfv}, IR resummation to describe 
the non-linear evolution of baryon acoustic oscillations~\cite{Blas:2016sfa,Ivanov:2018gjr},
and grid-based calculations for the matter bispectrum~\cite{Taruya_2021}.
} 
The two main goals of our work are (a) to define the validity 
range of this model and to (b) assess the information content 
of the redshift-space galaxy bispectrum in the tree-level 
approximation. Achieving these goals will bring us one step closer to 
understanding 
the information content of the galaxy 
bispectrum
and building a pipeline that can be used to analyze real data.

The paper is structured as follows. We describe the PT 
challenge simulations in Section~\ref{sec:data}.
Section~\ref{sec:theory} describes in detail our theoretical 
model. In Section~\ref{sec:like}
we discuss our baseline power spectrum and bispectrum
likelihoods. Our main results are 
presented in Section~\ref{sec:results},
where we analyze the real space and redshift space
monopole bispectrum data in combination with the baseline
redshift space power spectrum likelihood. 
We discuss improvements 
in cosmological and bias parameters
and give a forecast for a BOSS-like survey.
There we also compare the measured values of 
galaxy bias parameters with those expected from 
dark matter halo relations. 
We compare our analysis with previous works 
in Section~\ref{sec:compar}
and draw conclusions in Section~\ref{sec:concl}.
Several appendices contain additional material
and tests. In Appendix~\ref{app:bins} we validate our
binning approach, and in Appendix~\ref{app:coll}
we show that ``open'' triangles do not carry any 
significant cosmological information.
In Appendix~\ref{app:covar} we test our 
covariance matrix choices. Our baseline power 
spectrum likelihood is described in Appendix~\ref{app:power}. 
Appendix~\ref{app:real} contains an analysis 
of the power spectrum and bispectrum purely 
in real space.
Theoretical calculations of the power 
spectrum and bispectrum covariance matrices 
in perturbation theory are presented in Appendix~\ref{app:acovar}, while Appendix~\ref{app:fog}
contains a derivation of the Gaussian 
fingers-of-God damping.

\section{Data}
\label{sec:data}

The PT challenge simulation suite consists of 10 boxes, 
each with the side length $L=3840~h^{-1}$Mpc. The gravitational evolution was traced by 3072$^3$ particles in each box. In this paper we consider one particular snapshot taken at $z=0.61$,
which corresponds to the BOSS CMASS1 sample~\cite{Alam:2016hwk}. The dark matter halos from this snapshot
were populated with mock 
CMASS-like red luminous galaxies 
following the halo occupation distribution (HOD) prescription 
detailed in Ref.~\cite{Nishimichi:2020tvu}.
We refer the reader to this reference for further details on the simulations. The redshift-space 
power spectrum multipoles $\ell=0,2,4$ were estimated 
as 
\be
\label{eq:pest}
\hat P_\ell (k_i)=\frac{2\ell+1}{N_i}\sum_{\tilde \k\in k_i}\mathcal{L}_\ell(\mu_{\tilde \k})
\hat P (\tilde \k)\,,
\ee
where we have introduced
\be
 \hat P (\tilde \k)=\frac{\tilde V|\delta_{\tilde \k}|^2
 -\bar n^{-1}}{W^2_{\rm CIC}(\tilde \k)}\,,\quad 
 \tilde V=\left(\frac{D_A^{(\rm fid)}(z)}{D_A^{(\rm true)}(z)}\right)^2 \frac{H^{(\rm true)}(z)}{H^{(\rm fid)}(z)}L^3\,,
\ee
and $\delta_{\tilde \k}$ is the Fourier space 
overdensity field, the sum runs over all modes whose norms
belong to a  
 bin $[(i-1)\Delta k, i\Delta k]$,
 we use $\Delta k=0.01~\hMpc$ and
$N_i$
is the number of Fourier modes in the bin.
The modes in the  sum in Eq.~\eqref{eq:pest} 
are composed of fundamental modes $k_f=2\pi L^{-1}$,
which were rescaled 
by the 
Alcock-Paczynski (AP) effect~\cite{Alcock:1979mp} as,
\be
\tilde k_{f,x}= k_{f,x} \frac{D_A^{(\rm true)}(z)}{D_A^{(\rm fid)}(z)}\,,\quad 
\tilde k_{f,y}= k_{f,y} \frac{D_A^{(\rm true)}(z)}{D_A^{(\rm fid)}(z)}\,,\quad 
\tilde k_{f,z}= k_{f,z} \frac{H^{(\rm fid)}(z)}{H^{(\rm true)}(z)}\,,
\ee
where the upper scripts (true) and (fid) denote
the comoving angular diameter distance $D_A(z)$ and
the Hubble parameter $H(z)$
calculated in the true and fiducial cosmologies, respectively.
The fiducial cosmological model is the same as in Ref.~\cite{Nishimichi:2020tvu}, flat $\L$CDM with $\Omega_m^{(\rm fid)}=0.3$.
Note that we have subtracted the Poissonian 
shot noise power spectrum contribution 
\be
\frac{1}{\bar n}=\frac{L^3}{N_{\rm gal}}\,,
\ee
where $N_{\rm gal}$ is the total number of galaxies,
taking into account the interlacing technique for the aliasing correction and the CIC window function.
If not stated otherwise, we will be using the 
datavector $[P_0,P_2,P_4]$
with $\kmax=0.14~\hMpc$. In addition, we 
employ the transverse moment $Q_0$ (equivalent of the real space power spectrum), which is estimated from the redshift space multipoles via
\be
\hat  Q_0 = \hat P_0 -\frac{1}{2}\hat P_2 +\frac{3}{8}\hat P_4\,, 
\ee
see Refs.~\cite{Ivanov:2021haa} for more detail and also Refs.~\cite{Hamilton:2000du,Tegmark:2003uf,Scoccimarro:2004tg} for earlier works.
We use $Q_0$
in the range of scales 
\[
0.14~\hMpc \leq k<0.4~\hMpc\,,
\]
so that it is not correlated with the 
multipoles' datavector.

The angle-averaged (monopole) bispectrum is computed using the following estimator:
\be
\label{eq:est}
\hat B_{0} (k_1,k_2,k_3)=
\frac{1}{N_B(k_1,k_2,k_3)}
\sum_{\tilde \q_1\in k_1}
\sum_{\tilde \q_2\in k_2}
\sum_{\tilde \q_3\in k_3}
\delta_{\rm K}(\tilde\q_{123})
\delta_{\tilde \q_1}\delta_{\tilde \q_2}\delta_{\tilde\q_3}
\,,
\ee
where $\tilde \q_{123}\equiv \tilde\q_1+\tilde\q_2+
\tilde\q_3$,
and $\delta_{\rm K}(\q_{123})$ 
denotes the Kronecker delta function, 
\be
  \delta_{K}(\q_{123})=\begin{cases}
    1, & \text{if $\q_{123}=0$}.\\
    0, & \text{otherwise},
  \end{cases} 
\ee
and $N_B(k_1,k_2,k_3)$ is the number of fundamental triangles in the bin defined by wavenumber centers $(k_1,k_2,k_3)$. Each bin has width $\Delta k=0.01~\hMpc$, which is the same as for the power spectrum estimator.
Note that unlike the power spectrum, 
we do not subtract the shot noise contributions 
from the bispectrum.
The estimator Eq.~\eqref{eq:est}
is evaluated with FFTs using
the Scoccimarro method~\cite{Scoccimarro:2015bla}.

The real space bispectrum is calculated 
using the same formula Eq.~\eqref{eq:est},
but with the real space density $\delta^{\rm real}_\k$
and without the AP effect.

\section{Theory Model}
\label{sec:theory}

Let us describe our theoretical 
model for the redshift space bispectrum. 
We will discuss each relevant 
component separately.

In what follows we will work 
in the plane-parallel approximation. 
The galaxy density contrast field in redshift space
at quadratic order in perturbation theory
reads~\cite{Perko:2016puo,Desjacques:2018pfv}
\be
\begin{split}
\delta^{(z)}(\k)=& Z_1(\k)\delta^{(1)}(\k)+[Z_2(\delta^{(1)})^2]_\k+d_1\epsilon(\k)\,, \\
&+  d_2 b_1 \delta^{(1)}(\k) \epsilon(\k) + (d_1\epsilon(\k))f\mu^2 \theta^{(1)}(\k)  
+ ...\,,
\end{split}
\ee
where $\delta^{(1)},\theta^{(1)}$ are linear matter density  
and velocity divergence fields (satisfying $\theta^{(1)}=\delta^{(1)}$); 
$\epsilon$ is the stochastic galaxy overdensity 
field, $d_1,d_2$ are free parameters, and
the standard perturbation theory~\cite{Bernardeau:2001qr} kernels are 
given by:
\bseq 
\begin{align}
&Z_1(\k)  = b_1+f\mu^2\,,\\
&Z_2(\k_1,\k_2)  =\frac{b_2}{2}+b_{\mathcal{G}_2}\left(\frac{(\k_1\cdot \k_2)^2}{k_1^2k_2^2}-1\right)
+b_1 F_2(\k_1,\k_2)+f\mu^2 G_2(\k_1,\k_2)\notag\\
&\qquad\qquad\quad~~+\frac{f\mu k}{2}\left(\frac{\mu_1}{k_1}(b_1+f\mu_2^2)+
\frac{\mu_2}{k_2}(b_1+f\mu_1^2)
\right)
\,,\\
& F_2(\k_1,\k_2)=\frac{5}{7}
+\frac{1}{2}\left(
\frac{(\k_1\cdot \k_2)}{k_1^2}
+\frac{(\k_1\cdot \k_2)}{k_2^2}
\right)+\frac{2}{7}\frac{(\k_1\cdot \k_2)^2}{k_1^2 k_2^2}\,,\\
& G_2(\k_1,\k_2)=\frac{3}{7}
+\frac{1}{2}\left(
\frac{(\k_1\cdot \k_2)}{k_1^2}
+\frac{(\k_1\cdot \k_2)}{k_2^2}
\right)+\frac{4}{7}\frac{(\k_1\cdot \k_2)^2}{k_1^2 k_2^2}\,,
\end{align} 
\eseq
where $\mu_i\equiv (\k_i\cdot \hat{\z})/k_i$, $\mu\equiv (\k\cdot \hat{\z})/k$, $\k\equiv\k_1+\k_2$,
and $f$ is the logarithmic growth factor, related to the usual 
linear growth rate $D_+$
via 
\be
f=\frac{d\ln D_+}{d\ln a}\,,
\ee
with $a$ being the scale factor in the Friedmann metric.
The coefficients $b_1$, $b_2$, and $b_{\mathcal{G}_2}$
capture linear, quadratic, and tidal bias between 
matter and galaxies, respectively. 
The tree-level bispectrum 
is obtained by computing the three-point function 
of the perturbative density field at second order~\cite{Desjacques:2016bnm}, 
\be
\begin{split}
& B_{\rm ggg}(\k_1,\k_2,\k_3)=
~2Z_2(\k_1,\k_2)Z_1(\k_1)Z_1(\k_2)
% P^{\rm IR -res}_{\rm tree}( \k_1) 
% P^{\rm IR -res}_{\rm tree}( \k_2)
P_{\rm lin}(k_1) 
P_{\rm lin}(k_2)
\\
&+
P_{\epsilon}(k_2)
% \frac{
% B_{\rm shot}
% }{\bar n}
% +\frac{B_{\rm shot}}{\bar n}  
2d_1 \left(d_2 b_1+
d_1 
f\mu^2_1
\right)Z_1(\k_1)P_{\rm lin}(k_1) 
+\text{cycl.} + 
d_1^3B_{\epsilon}(\k_1,\k_2,\k_3) \,,
\end{split}
\ee
where we have used the following correlation functions
\be
\begin{split}
& \langle \delta^{(1)}(\k)\delta^{(1)}(\k')\rangle  
=(2\pi)^3 \delta^{(3)}_D(\k+\k')P_{\rm lin}(k)\,,
\\
&
\langle \epsilon(\k)\epsilon(\k')\rangle  
=(2\pi)^3 \delta^{(3)}_D(\k+\k')P_{\epsilon }(k)\\
& \langle \epsilon(\k_1)
\epsilon(\k_2)
\epsilon(\k_3)
\rangle  = (2\pi)^3 
\delta^{(3)}_D(\k_1+\k_2+\k_3)B_\epsilon(\k_1,\k_2,\k_3)\,.
\end{split}
\ee
\paragraph{Stochastic terms.} At quadratic order in 
perturbation theory
the shot-noise contributions are constants,
\be
P_\epsilon =\text{const}\,, \quad B_\epsilon =\text{const}\,.
\ee
Furthermore, if $\epsilon$ is Poisson-distributed, both statistics are fully 
determined by the galaxy number density $\bar n$ (see e.g.~Ref.~\cite{Schmittfull:2014tca} and references therein):
\be
\label{eq:BPeps}
B_\epsilon= P_\epsilon^2 =\frac{1}{\bar n^2}\,.
\ee
However, due to halo exclusion, deviations from Poissonian sampling are known 
to be important~\cite{Casas-Miranda:2001dwz,Baldauf:2013hka,Baldauf:2015fbu,Schmittfull:2018yuk}, in which case we cannot use Eq.~\eqref{eq:BPeps}
and the tree-level bispectrum should be characterized by 
three free parameters capturing stochasticity. We define them 
to be $P_{\rm shot}, B_{\rm shot}$ and $A_{\rm shot}$;
\be
\begin{split}
d_1^2 \langle \epsilon^2 \rangle = \frac{1+P_{\rm shot}}{\bar n}\,,\quad 
d_1^3 \langle \epsilon^3 \rangle = \frac{A_{\rm shot}}{\bar n^2 }\,,\quad B_{\rm shot} \equiv 2 d_2 d^{-1}_1\left(1+P_{\rm shot}\right)\,,
\end{split} 
\ee
which are expected to be $\mathcal{O}(1)$ numbers.
Importantly, the parameter $P_{\rm shot}$ also enters 
the power spectrum model. Furthermore, following~\cite{Angulo:2015eqa,Gil-Marin:2014sta,DAmico:2019fhj} we will assume
that the bispectrum and power spectrum of the stochastic overdensity 
component are correlated as in the Poissonian case~\eqref{eq:BPeps},
but their values are different from $\bar n^{-1}$, i.e.
\be 
B_\epsilon= P_\epsilon^2 \,,\quad \Rightarrow \quad 
A_{\rm shot}=(1+P_{\rm shot})^2\,,
\ee 
which is ultimately motivated by the expectation 
that departures from the Poissonian sampling are small.
We have found 
that the bispectrum data is fully consistent 
with this hypothesis. 
Therefore, we adopt 
this choice as our baseline model for the stochastic nuisance parameters, 
which helps us reduce their number
down to two.

\paragraph{Fingers-of-God.}
An important feature of non-linear redshift-space distortions
is the sensitivity to the stochastic velocity 
field, which can have relatively 
large scale correlations due to halo virialization~\cite{Scoccimarro:2004tg}.
This effect is called ``fingers-of-God'' (FoG)~\cite{Jackson:2008yv}.
In the EFT, FoG are captured perturbatively 
through the 
gradient expansion involving derivatives 
along the 
line-of-sight~\cite{Senatore:2014vja,Lewandowski:2015ziq,Perko:2016puo,Vlah:2018ygt,Chen:2020fxs,Chen:2020zjt}. 
These corrections are called
``counterterms,'' and at leading one-loop order they are given by~\cite{Perko:2016puo}
\be 
\label{eq:ctrs}
\delta^{\rm ctr.}=
-c_0\left(\frac{k}{k_{\rm NL}}\right)^2-(c_1\mu^2 +c_2 \mu^4)
\left(\frac{k}{k^r_{\rm NL}}\right)^2\,.
% \quad k^r_{\rm NL}=0.3~\hMpc\,.
% K^2=\sum_{a=1}^3 k_a^2 \mu_a^2\,,\quad \mu_a\equiv (\hat \k_a \cdot \hat \z)\,,
\ee
The role of $\mu$-dependent counterterm coefficients $c_1$
and $c_2$
is to capture the physical impact of the FoG
on large scale fluctuations.\footnote{Strictly speaking, each coefficient $c_i$ has ``infinite'' and ``finite''
pieces. 
The role of the infinite piece is to renormalize 
the UV part of one-loop integrals, whilst the ``finite''
part captures physical backreaction from short scales. }
In principle, the FoG
is a one-loop effect in the EFT nomenclature,
and it needs to be included along with other, ``standard,'' one-loop 
corrections, which we ignore in this work.
The characteristic momentum scale of 
these one-loop
corrections matches
the real space dark matter cutoff\footnote{The EFT calculations, at least at the one loop order, can be interpreted as so-called ``standard perturbation theory''~\cite{Bernardeau:2001qr} computations corrected with 
a set of UV ``counterterms.'' In this picture the one-loop
integrals have the same scaling for all tracers, while the tracer-specific momentum cutoffs appears only from 
the counterterms.} $k_{\rm NL}$.
If $2\pi/k^{r}_{\rm NL}$
is larger than $2\pi k^{-1}_{\rm NL}$ (and the 
cutoff of the bias expansion $2\pi k^{-1}_{\rm M}$),
the FoG counterterm 
can actually dominate over usual 
loop corrections. 
This is the exact situation that was observed
for matter and galaxy power spectra in redshift space, where 
FoG corrections
were found to be important even on relatively large scales
where the ``standard'' loop corrections (i.e.~without the counterterms) are suppressed~\cite{Lewandowski:2015ziq,Ivanov:2019pdj,Nishimichi:2020tvu,Chudaykin:2020hbf,Ivanov:2021zmi}.\footnote{Note that the form of the finite counterterms in Eq.~\eqref{eq:ctrs} is quite similar to the large-scale limit of some 
phenomenological prescriptions for FoG, e.g.~the Gaussian damping model, see Appendix~\ref{app:fog} for more detail.}

This motivates including the 
FoG counterterms $c_1,c_2$ in our theory model even though
formally they capture one-loop effects. Another rationale 
behind this practice is that these counterterms can be treated 
as a proxy for the theoretical error~\cite{Baldauf:2016sjb,Chudaykin:2020hbf}. This will
also 
serve 
us as a tool to check if the tree-level 
calculation 
can be trusted: 
if the counterterm contribution 
dominates the
tree-level bispectrum signal, the one-loop corrections
cannot be ignored anymore. 

In practice, we have found that it is sufficient to include only the $k^2 \mu^2$
counterterm in our theory model. We ignore the contribution $k^2 \mu^4$
because we have found 
that it is very degenerate with the $k^2 \mu^2$ shape 
at the level of the bispectrum monopole, and hence we
set $c_2=0$ in what follows. Note that we will have to include
both $c_1$ and $c_2$ when we consider 
higher order angular multipole moments.
The inclusion of the $c_1$ counterterm amounts to 
correcting the kernel $Z_1$ as
\be
Z_1\to  Z^{\rm FoG}_1 = b_1 + f\mu^2 -c_1\mu^2 \left(\frac{k}{k^r_{\rm NL}}\right)^2\,.
\ee 
In what follows we set $k_{\rm NL}^r=0.3~\hMpc$
in agreement with the measurement of the cutoff for the Red Luminous Galaxies
from the power spectrum of the PT challenge mocks~\cite{Nishimichi:2020tvu,Ivanov:2021zmi}.

\paragraph{IR resummation.} 

Naive attempts to build the EFT as a 
perturbative expansion in terms of smoothed (large-scale)
density and velocity fields break down for the 
BAO part of the linear power spectrum (sometimes loosely referred to as the ``BAO wiggles'').
The procedure of resumming enhanced perturbative (loop) 
corrections to this part of the spectrum 
is called ``IR resummation''~\cite{Senatore:2014via,Baldauf:2015xfa,Blas:2015qsi,Vlah:2015sea,Vlah:2015zda,Blas:2016sfa} (see Refs.~\cite{Crocce:2007dt,Crocce:2007dt} for earlier works).
IR resummation effects have to be included
in the theory model even when it is evaluated at the
tree level~\cite{Baldauf:2015xfa,Blas:2016sfa}.
IR-resummation 
for the bispectrum in redshift space has 
been calculated in Ref.~\cite{Ivanov:2018gjr} (see Ref.~\cite{Vasudevan:2019ewf} for IR resummation of the bispectrum in the case of non-Gaussian initial 
conditions). 
At leading order 
this procedure amounts to the replacement 
of the linear matter power spectrum by its 
resummed version,
\be
\label{eq:IRres}
P_{\rm lin}(k)\to P^{\rm IR -res}_{\rm tree}(k)= P_{\rm nw}(k)+ 
P_{\rm w}(k)\e^{-\Sigma^2k^2 (1+f\mu^2(2+f))-\delta \Sigma^2 k^2 f^2 \mu^2(\mu^2-1)}\,,
\ee
where $P_{\rm w}$ is the part of the spectrum that
contains the BAO wiggles, $P_{\rm nw}\equiv P_{\rm lin}-P_{\rm w}$, 
\be 
\begin{split}
& \Sigma^2 =\frac{1}{6\pi^2}\int_0^{k_S} dq~P_{\rm nw}(q)
(1-j_0(qr_{\rm BAO})+2j_2(qr_{\rm BAO}))\,,\\
& \delta\Sigma^2 = \frac{1}{2\pi^2}\int_0^{k_S} dq~P_{\rm nw}(q)
j_2(qr_{\rm BAO})\,,
\end{split}
\ee
are the BAO damping functions, 
$j_\ell(x)$ are spherical Bessel functions,
$r_{\rm BAO}$ is the comoving 
sound horizon at the drag epoch,
$k_S$ is the separation scale
defining IR modes that need to be resummed. In practice we use 
$k_S=0.05~h/$Mpc following Ref.~\cite{Blas:2016sfa}, 
although other choices, e.g.~$k_S=k/2$~\cite{Baldauf:2015xfa}
give statistically indistinguishable results.

All in all, our final 
tree-level bispectrum model reads (c.f.~\cite{Desjacques:2018pfv}):
\be
\label{eq:Bfull}
\begin{split}
B_{\rm g g g}=
& \left[~2Z_2(\k_1,\k_2)Z^{\rm FoG}_1(\k_1)Z^{\rm FoG}_1(\k_2)
P^{\rm IR -res}_{\rm tree}( \k_1) 
P^{\rm IR -res}_{\rm tree}( \k_2)\right.\\
&\left.+\frac{B_{\rm shot}}{\bar n}
% +\frac{B_{\rm shot}}{\bar n}  
\left( b_1+2\frac{1+P_{\rm shot}}{B_{\rm shot}}f\mu^2
\right)Z^{\rm FoG}_1(\k_1)P^{\rm IR -res}_{\rm tree}(\k_1) 
+\text{cycl.}\right] + \frac{(1+P_{\rm shot})^2}{\bar n^2} \,,
\end{split}
\ee
where $b_1$, $b_2$, $b_{\mathcal{G}_2}$, $P_{\rm shot}, 
B_{\rm shot}$ are nuisance parameters to marginalize over.
Note that $B_{\rm shot}$ is the only new parameter that is not
present in the power spectrum model.

\paragraph{Redshift space multipoles.}
In real space the bispectrum depends on three kinematic variables (wavelengths) 
which characterize
the shape of a triangle.
In redshift space there appears an additional dependence due to 
the orientation of the triangle w.r.t.~the line-of-sight direction.
This orientation is characterized by two angles,
which we choose, following Ref.~\cite{Scoccimarro:1999ed}, to be the polar angle of $\k_1$ (its cosine is $\cos\theta = \mu\equiv (\hat{\k}_1\cdot \hat{\z})$) and the azimuthal angle around $\k_1$
denoted by $\phi$. In this case the angles between 
wavevectors $\k_a$ (a=1,2,3) and the line-of-sight are given by  
\be
\label{eq:angles}
\begin{split}
& \mu_1 = \mu \,,\\
&  \mu_2 = \mu \cos\alpha -(1-\mu^2)^{1/2}\sin \alpha \cos \phi\,,\\
& \mu_3 = -\frac{k_1}{k_3}\mu -\frac{k_2}{k_3}\mu_2 \,,
\end{split}
\ee
where $\cos \alpha = x=(\hat{\k}_1\cdot \hat{\k}_2)$. It is convenient 
to describe this angular dependence by expanding $B_{\rm ggg}$ in spherical harmonics,
\be
\begin{split}
& B_{\rm ggg}(\k_1,\k_2,\k_3)=\sum_{\ell =0}^\infty\sum_{m=-\ell}^{\ell}B_{\ell m}(k_1,k_2,k_3)Y_{\ell m}(\theta,\phi) \,,\\
& B_{\ell m}(k_1,k_2,k_3)=\frac{2\ell+1}{2}\int_{0}^{2\pi} d\phi \int_{-1}^{1}d(\cos\theta)~ Y^*_{\ell m}(\theta,\phi)~B_{\rm ggg}(\k_1,\k_2,\k_3)\,.
\end{split}
\ee
In what follows we will focus on the $m=0$ sector~\cite{Scoccimarro:2015bla}. The corresponding momenta $B_\ell$ are called ``bispectrum multipoles,''
\be
\label{eq:Bell}
\begin{split}
B_{\ell}(k_1,k_2,k_3)=\frac{2\ell+1}{2}\int_{0}^{2\pi} \frac{d\phi}{2\pi} \int_{-1}^{1}d(\cos\theta)~ \mathcal{L}_{\ell}(\cos\theta)~B_{\rm ggg}(\k_1,\k_2,\k_3)\,,
\end{split}
\ee
where $\mathcal{L}_{\ell}$ denotes a 
Legendre polynomial of order $\ell$.
Note that the integral above can be done analytically 
at the tree level in the absence of 
IR resummation and the AP effect~\cite{Scoccimarro:1999ed}. However, in what follows we will use 
the full formula \eqref{eq:Bfull} with IR resummation,
and evaluate angular integrals in Eq.~\eqref{eq:Bell}
numerically via Gauss-Legendre quadrature. 

\paragraph{Alcock-Paczynski effect.}
The AP conversion~\cite{Alcock:1979mp} from true wavenumbers and angles ($q,\nu$)
to observed wavenumbers and angles $(k,\mu)$ is given by
\be
\begin{split}\label{AP_k_mu}
	q^2&=k^2\left[
	% \l
	\a_\parallel^{-2}
	% \frac{H_\true}{H_\fid}
	% \r^2
	\mu^2+
	\a_\perp^{-2}
	(1-\mu^2)\right]\,,\\
	\nu^2&=
	\a_\parallel^{-2}
	% \l\frac{H_\true}{H_\fid}\r^2
	\mu^2\left[
	\a_\parallel^{-2}
	% \l\frac{H_\true}{H_\fid}\r^2
	\mu^2+
	\a_\perp^{-2}
	% \l\frac{D_{A,\fid}}{D_{A,\true}}\r^2
	(1-\mu^2)\right]^{-1}\,,
\end{split}
\ee
which depends on 
the ratios between the true and fiducial Hubble parameters 
and angular diameter distances at the redshift of interest,
\be
\alpha_\parallel = \frac{H_{\rm fid}(z)}{H_{\rm true}(z)}
\frac{H_{0,\rm true}}{H_{0,\rm fid}}
\,,\quad \alpha_\perp = \frac{D_{\rm true, A}(z)}{D_{\rm fid, A}(z)}
\frac{H_{0,\rm true}}{H_{0,\rm fid}}\,,
\ee
where additional factors $H_{0,\rm true}/H_{0,\rm fid}$ account for the fact 
that wavenumbers are measured in $\hMpc$ units in our analysis.
The observed power spectrum multipoles are given by~\cite{Chudaykin:2020aoj} 
\be
P_\ell(k)=\frac{2\ell+1}{2\alpha^2_\perp \alpha_\parallel}\int_{-1}^1 d\mu~\mathcal{L}_\ell(\mu) P_{\rm gg}(q[k,\mu],\nu[\mu])\,.
\ee
In full analogy, the bispectrum multipoles are given by~\cite{Song:2015gca}
\be
\begin{split}
&B_{\ell}(k_1,k_2,k_3)\\
&=\frac{2\ell+1}{2\alpha^2_\parallel\alpha^4_\perp}
\int_{0}^{2\pi} \frac{d\phi}{2\pi} \int_{-1}^{1}d\mu_1~ 
\mathcal{L}_{\ell}(\mu_1)~B_{\rm ggg}(q_1[k_1,\mu_1],q_2[...],q_3[...],\nu_1[\mu_1],\nu_2[\mu_2(\mu_1)])
% [\k^{\obs}_1,\k^{\obs}_2,\k^{\obs}_3]
\,,
\end{split}
\ee
where the observed angles satisfy Eq.~\eqref{eq:angles}. In what follows we will
focus on the monopole moment $\ell=0$, and leave the analysis 
of other multipoles for future work.

\paragraph{Binning effects.}
The measured bispectrum is a discrete approximation to a continuous Fourier-space field. 
In order to account for this discreteness we need to bin our 
theory predictions in the same way as we bin the data.
Binning corrections are marginally important 
for the PT challenge power spectrum
and it is straightforward 
to take them into account~\cite{Nishimichi:2020tvu}.
However, the situation is somewhat different for the bispectrum,
where binning can be a serious source of systematics~\cite{Bernardeau:2011dp}.
The exact discrete bispectrum that we extract from simulations is given by\footnote{We omit the subscript `${\rm ggg}$' for clarity in this section, i.e.~replace $B_{\rm ggg}\to B$.} 
\be
\label{eq:Bdisc}
\begin{split} 
& \hat{B}_{0,\rm disc}(k_1,k_2,k_3)=\frac{\sum_{\q_1\in k_1} \sum_{\q_2\in k_2} \sum_{\q_3\in k_3}
B(\q_1,\q_2,\q_3)
\delta_{\rm K}(\q_{123})}{N^{\rm disc}_B(k_1,k_2,k_3)}\,,\\
& N^{\rm disc}_B(k_1,k_2,k_3) = \sum_{\q_1\in k_1} \sum_{\q_2\in k_2} 
\sum_{\q_3\in k_3}\delta_{\rm K}(\q_{123})\,.
\end{split}
\ee
The sum in Eq.~\eqref{eq:Bdisc} runs over all
discrete wavevectors $\q_i$ that belong to the triangle bin 
defined by its center $(k_1,k_2,k_3)$
and width $\Delta k$. 
$N^{\rm disc}_B(k_1,k_2,k_3)$ is the total number 
of these ``fundamental triangles'' inside the triangle 
bin $(k_1,k_2,k_3)$~\cite{Oddo:2019run}.

Before going into technical details, let us outline 
our strategy.  
As a first step, we take the
continuum limit, i.e.~assume a vanishingly small 
fundamental wavenumber $k_f=(2\pi)L^{-1}$
as a leading approximation. 
In this first approximation the discreetness effects 
can be taken into account by integrating the continuous 
bispectrum field within appropriate bins. 
It is natural to refer to this program as the 
``integral approximation.''
Because the actual fundamental bin is finite, 
the integral approximation requires certain 
corrections.
As a second step, we will introduce these corrections,
which will be referred to as ``discreetness weights.''

Note that with our binning scheme there are so-called 
``open'' triangle bins. The centers of these bins 
$(k_1,k_2,k_3)$ do not satisfy 
momentum conservation constraints, such as $|k_3-k_2|<k_1<k_3+k_2$.\footnote{Individual triangles
that belong to the bin are, of course, valid 
triangles that satisfy all relevant constraints.}
In what follows we will 
discard these triangles because of three reasons: 
\begin{itemize}
\item their properties (and very existence) crucially depend on the box size, which 
makes it hard to make generic statements that would not depend 
on a particular survey volume;
\item the leading binning effect 
cannot be well captured by 
the 
integral approximation for
these triangles, and hence it requires 
a significant modification of our baseline binning program;
\item these triangles do not carry any sizable cosmological 
information (at least at the level of the tree-level bispectrum likelihood) and with our particular choice of bins' width), see Appendix~\ref{app:coll}.
\end{itemize}
As a first step of our binning procedure we implement
the integral approximation.
Other binning schemes were 
explored in Refs.~\cite{2010MNRAS.406.1014S,Oddo:2019run,Eggemeier:2021cam}.
The integral approximation
amounts to replacing the sum over the modes 
with the Fourier integral,
\be
\begin{split}
& \sum_{\q_1\in k_1} \sum_{\q_2\in k_2} 
\sum_{\q_1\in k_2}
\delta_{\rm K}(\q_{123})
B(\q_1,\q_2,\q_3)
\to  
\frac{V^2}{(2\pi)^6}\int_{\k_1\k_2\k_3} 
~(2\pi)^3\delta^{(3)}_D(\q_{123})
B(\q_1,\q_2,\q_3)\,,\\
& \sum_{\q_1\in k_1} \sum_{\q_2\in k_2} \sum_{\q_1\in k_2}
\delta_{\rm K}(\q_{123})\to  
\frac{V^2}{(2\pi)^6}\int_{\k_1\k_2\k_3} 
% [d^3q]^3
~(2\pi)^3\delta^{(3)}_D(\q_{123})
\end{split}
\ee
where $V$ is the box volume and we introduced
\be 
\label{eq:intBN}
\begin{split}
& \int_{\k_1\k_2\k_3} \equiv \int_{V_{k_1k_2k_3}}\frac{d^3q_1}{(2\pi)^3}
\frac{d^3q_2}{(2\pi)^3}
\frac{d^3q_3}{(2\pi)^3}\,,\quad  V_{k_1k_2k_3}=\mathcal{D}_1\times\mathcal{D}_2\times \mathcal{D}_3\,,\\
& \mathcal{D}_a = \left\{ (q_{x_1},q_{x_2},q_{x_3})\in \mathbb{R}^3:k_a-\frac{\Delta k}{2}\leq |\q_a|\leq k_a+\frac{\Delta k}{2}\right\}\,,\quad a=1,2,3\,.
\end{split}
\ee
\begin{figure}
\centering
\includegraphics[width=0.49\textwidth]{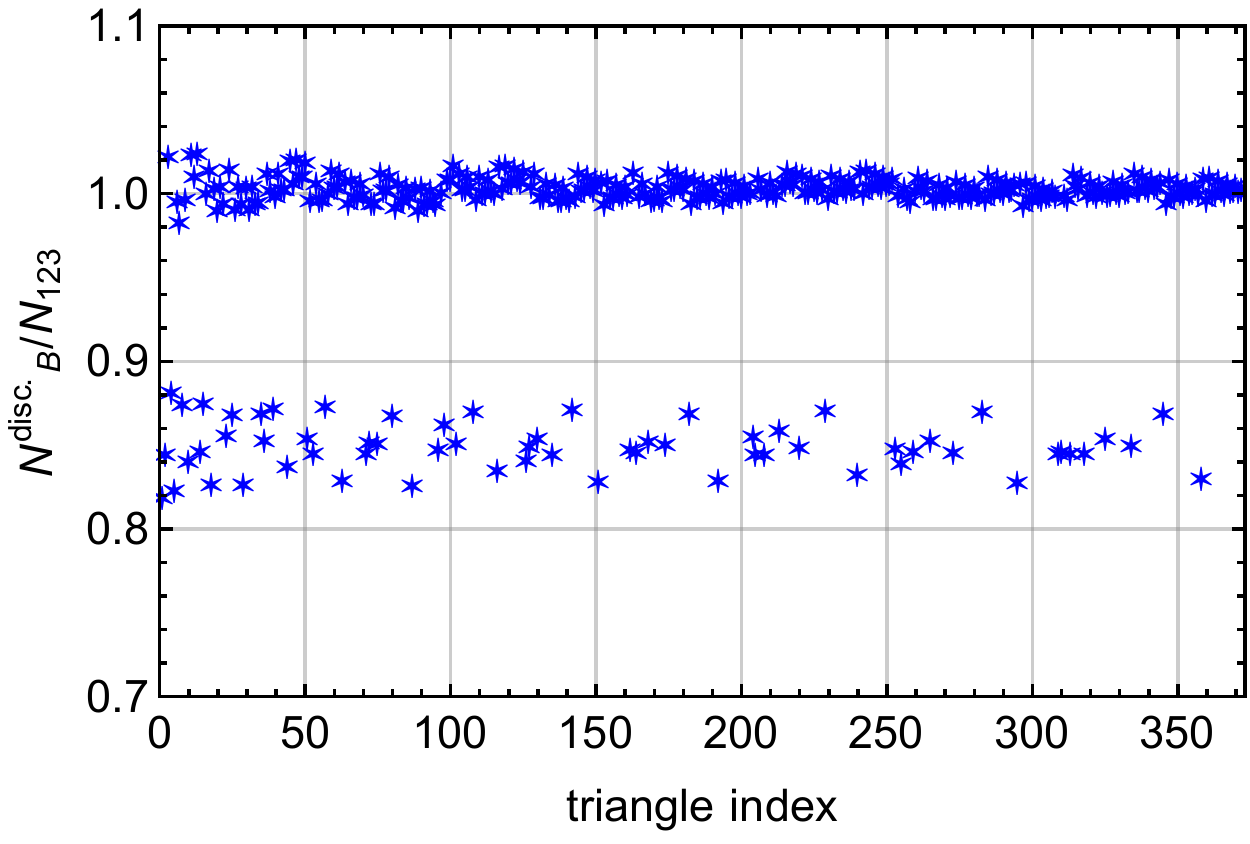}
\includegraphics[width=0.49\textwidth]{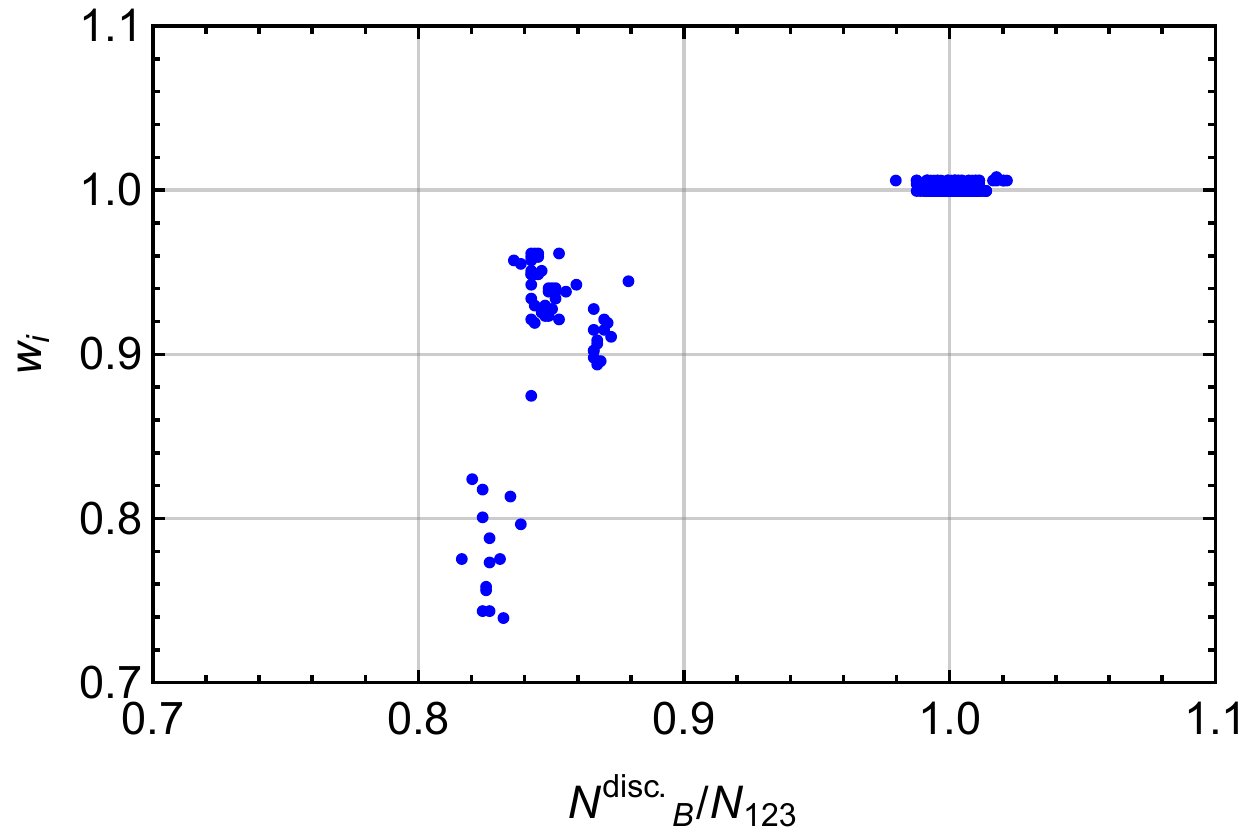}
 \includegraphics[width=0.65\textwidth]{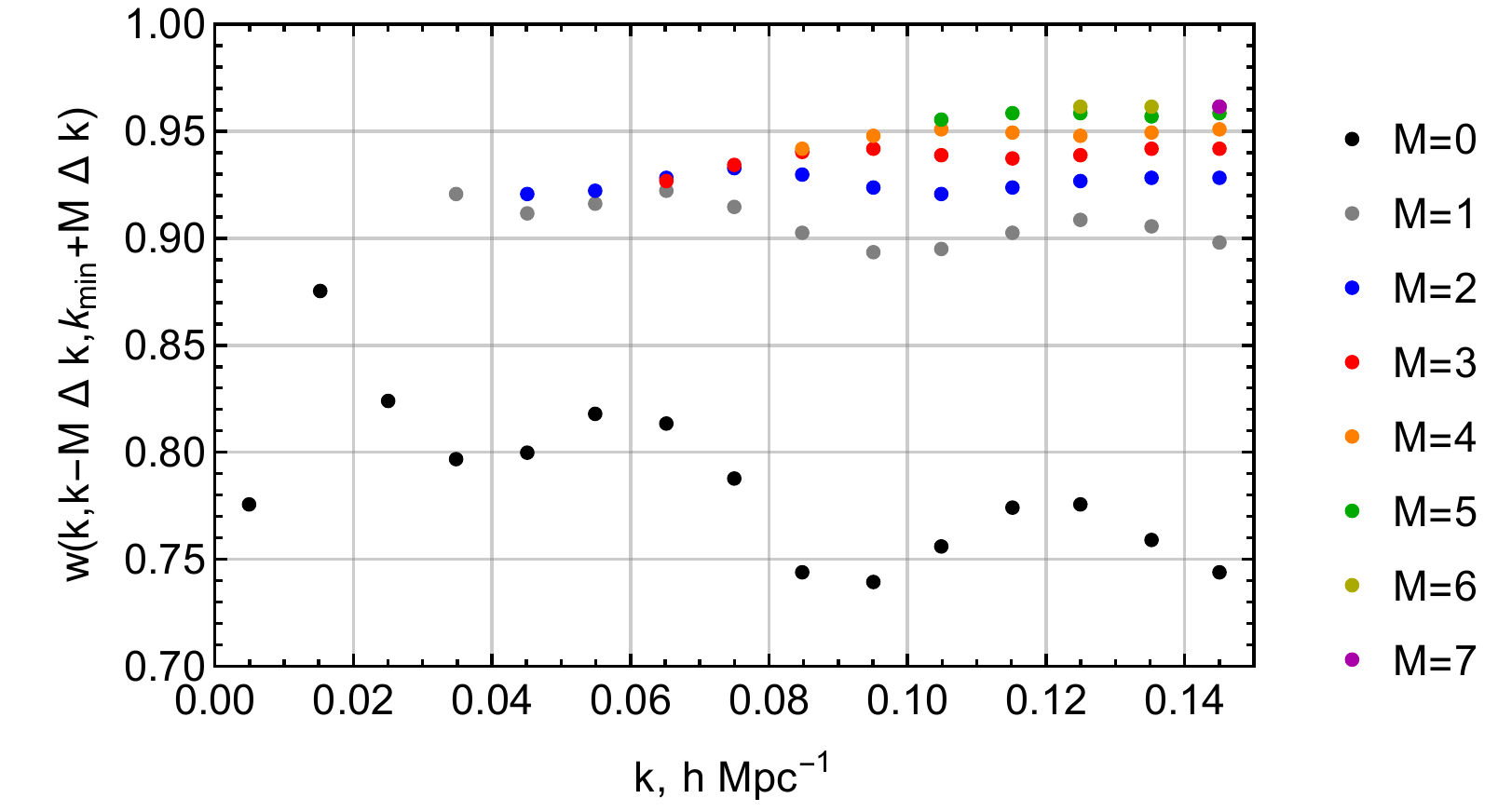}
\caption{\textit{Upper left panel:} the ratio 
between the number of fundamental triangles 
from the data 
and from the integral approximation, $N_{B}^{\rm disc.}/N_{123}$, as a function
of the triangle bin. 
\textit{Upper right panel:} 
the ratio 
between the exact binned bispectrum 
and the integral approximation 
(``discreteness weights'' $w_i$), as a function
of $N_{B}^{\rm disc.}/N_{123}$ for the same triangle bins. 
The triangles shown correspond to 
$k_{\rm max}=0.15~h/$Mpc.
\textit{Lower panel:} discreteness weights for folded triangles
denoted by their bin centers
(we use $\Delta k=0.01~\hMpc$, and hence $k_{\rm min}=\Delta k/2=5\cdot 10^{-3}~\hMpc$).
}
\label{fig:weights}
\end{figure}
This way we arrive at
\be
\label{eq:Bint}
\begin{split}
& \hat B_{0,\rm int}=
V^2
 \int_{\k_1\k_2\k_3} ~\frac{B(\q_1,\q_2,\q_3)}{N^T_{123}}
 (2\pi)^3\delta^{(3)}_D(\q_1+\q_2+\q_3)\,,\\
& N_{123}=V^2\int_{\k_1\k_2\k_3}~(2\pi)^3 
\delta^{(3)}_D(\q_{123})=8\pi k_1k_2k_3 \Delta k^3\frac{V^2}{(2\pi)^6}\,.
\end{split} 
\ee
The delta-function can be integrated explicitly 
following Ref.~\cite{Mehrem:1990eg},
yielding
\be
\label{eq:Bint2}
\begin{split}
\hat B_{0,\rm int}=\frac{V^2}{(2\pi)^6N_{123}}
\int_0^{2\pi} \frac{d\phi}{4\pi} \int_{-1}^1d\mu
\left(\prod_{i=1}^3\int^{k_i+\Delta k/2}_{k_i-\Delta k/2} dq_i~q_i\right)
~B(q_1,q_2,q_3,\mu,\phi) 
\,.
\end{split} 
\ee 
In order to estimate the accuracy of 
the integral approximation, we compare 
the continuous ($k_f\to 0$ limit) prediction for the number 
of triangle modes that fall in a given bin $N_{123}$ 
with the actual number of discrete triangles 
in that bin $N_B^{\rm disc.}$.
The result is shown in the upper left panel of Fig.~\ref{fig:weights}, where we display the ratio $N_B^{\rm disc.}/N_{123}$ for the bins whose centers satisfy the momentum 
conservation constraint, and which we actually use 
in the analysis. We see that the integral approximation
correctly predicts the number of fundamental triangles
for most of the bins, up to a few percent precision. 
However the integral approximation is not 
very accurate for folded triangles with $k_2+k_3=k_1+\Delta k/2$.
For these triangles the typical mismatch is about $\sim 15\%$. 
This discrepancy also leads to a mismatch at the level of binned bispectra.
To correct for this discrepancy we introduce ``discreteness weights'' $w$,
\be
w = \frac{\hat B_{\rm disc}}{\hat B_{\rm int}} \,,
\ee
where $\hat B_{\rm disc}$ is computed by using 
a direct discrete expression~Eq.~\eqref{eq:Bdisc}, while 
$\hat B_{\rm int}$ is calculated from Eq.~\eqref{eq:Bint}.

We compute the weights for a certain fiducial cosmology 
and nuisance parameters extracted 
from a fit to the simulation data analyzed without the weights. 
Since the evaluation of the 
full discrete expression~Eq.~\eqref{eq:Bdisc}
is too expensive for an Markov chain Monte Carlo (MCMC), the best strategy would be 
to iterate the discreetness weights for 
the best-fit bispectra from a few consecutive 
MCMC runs. 
However, quite remarkably, we have found that this iterative 
procedure has converged already at the first step.
Our initial fiducial parameters happened to be 
significantly different from the actual best-fit parameters,
yet both produced almost identical discreteness weights. 
This shows that the discreetness weights are nearly cosmology-independent, hence they can be computed only once 
for a given  
survey specification.

We display the discreteness weights for PT challenge
boxes in the right panel of Fig.~\ref{fig:weights}, along with the ratio 
$N_B^{\rm disc.}/N_{123}$, which demonstrates that 
the ``problematic'' triangles can be easily identified in the data
by comparing the number of fundamental triangles in the bin 
with the prediction of the integral approximation. 
As we can see from this figure, these corrections need to be included if $N_B^{\rm disc.}/N_{123}$ deviates 
from unity by more than $10\%$.
We show discreteness weights specifically for the problematic folded triangles 
in the lower panel of Fig.~\ref{fig:weights}. For all other triangle configurations
the discreteness weights coincide with unity with $\mathcal{O}(0.5)\%$ precision, implying that the integral approximation is very accurate for them.

Additionally, we validate our discreteness weights 
approach in Appendix~\ref{app:bins} by comparing it with 
an approximate discrete binning scheme similar to Eq.~\eqref{eq:Bdisc}.
These tests suggest that our treatment of discreteness effects 
is accurate enough for the full simulation volume
and hence can be safely adopted for the purposes of our paper
and for any realistic future analysis.

All in all our theory model is given by  
\be 
B^{\rm th}=\hat B^{\rm int}_0(k_1,k_2,k_3)w(k_1,k_2,k_3)\,,
\ee
where $\hat B^{\rm int}_0$ is computed from Eq.~\eqref{eq:Bint2}
by numerically performing the five-dimensional integral over
the tree-level IR resummed model~\eqref{eq:Bfull}.

\section{Likelihood}
\label{sec:like}

We will use a Gaussian likelihood for the bispectrum~\cite{Sefusatti:2006pa},
\be
\ln \mathcal{L}_{\rm B}=- \frac{1}{2}\sum_{\text{triangles}~T'} 
(B^{\rm th}_T-B^{\rm data}_T) (B^{\rm th}_{T'}-B^{\rm data}_{T'}) 
(C^{\rm B})^{-1}_{TT'}\,,
\ee
where we assume without loss of generality that the bin centers
satisfy $k_1\geq k_2\geq k_3$ and
\be
\begin{split}
& \sum_{T}\equiv \sum_{k_1=k_{\rm min}}^{k_{\rm max}}
\sum_{k_2=k_{\rm min}}^{k_1}\sum_{k_3=k_*}^{k_2}\,,\quad  k_*\equiv \text{max}(k_{\rm min},k_1-k_2)\,.
\end{split}
\ee
The Gaussian likelihood approximation for the bispectrum
is justified within perturbation theory, which is consistent 
with the tree-level approximation for the bispectrum itself. 
This approximation must be true on sufficiently 
large scales,
to which we limit our analysis. 
In this regime we can use the Gaussian tree-level 
approximation for the covariance matrix $C^{\rm B}$~\cite{Scoccimarro:1999ed,Sefusatti:2006pa,Song:2015gca,Scoccimarro:2015bla},
\be
\begin{split}
& C^{\rm B}_{TT'}=\frac{(2\pi)^3\pi s_{123}}{k_1k_2k_3\Delta k^3 V_{\rm tot}}\delta_{TT'}
\int_0^{2\pi} \frac{d\phi}{4\pi}\int_{-1}^1 d\mu ~\prod_{i=1}^3 \left[
P_{\rm lin}(k_i)(b_1+f \mu^2_i(\phi,\mu))^2 +\frac{1}{\bar n}
\right]\,,
\end{split} 
\ee
where $(k_1,k_2,k_3)$ denotes the center of the triangle bin $T$, 
$V_{\rm tot}$ is the cumulative volume of the PT challenge simulations
($V_{\rm tot}=566~h^{-3}$Gpc$^3$),
$s_{123}=~$6, 2 or 1 for equilateral, isosceles and general triangles.
To approximately account for the discreteness binning effects we use the true number of fundamental
triangles in the bin instead of the prediction of the integral approximation,
i.e.~we rescale
\be 
C^{\rm B}_{TT'} \to \frac{N_{123}}{N^{\rm disc}_B} \cdot C^{\rm B}_{TT'} \,.
\ee
We evaluate the covariance for the best-fit cosmology extracted from 
the power spectrum likelihood analysis.
We ignore the cross-covariance between the power spectrum and the bispectrum 
in our baseline analysis.
This and other likelihood 
approximations
are validated 
in Appendix~\ref{app:covar}.
There we show
that our results are stable if we include the one-loop theoretical error
bispectrum 
covariance, and 
the 
cross-covariance between the power
spectrum and bispectrum (computed in perturbation theory), 
as well as if we replace
the Gaussian
bispectrum
covariance
with the sample
covariance 
from the available 
mocks. All these 
different 
options yield
statistically 
indistinguishable 
results. 

Our total likelihood thus consists of a product
of the bispectrum and baseline power spectrum likelihoods,
\be 
\mathcal{L}_{\rm tot}=\mathcal{L}_{\rm B}\times \mathcal{L}_{\rm P}\,.
\ee
The details of our baseline power spectrum likelihood can be 
found in Appendix~\ref{app:power}
and in Ref.~\cite{Ivanov:2021haa}. 
We compute power spectrum theoretical templates 
using the \text{CLASS-PT} code~\cite{Chudaykin:2020aoj}.\footnote{Publicly available at 
\href{https://github.com/Michalychforever/CLASS-PT}{
\textcolor{blue}{https://github.com/Michalychforever/CLASS-PT}}
}
We run MCMC chains 
using the \texttt{Montepython} code~\cite{Brinckmann:2018cvx,Audren:2012wb}.\footnote{Publicly available at 
\href{https://github.com/brinckmann/montepython\_public}{
\textcolor{blue}{https://github.com/brinckmann/montepython\_public}}
} 
Posterior density plots are generated with the 
\texttt{getdist} package~\cite{Lewis:2019xzd}.
We will scan over the parameters of the base $\Lambda$CDM model
and EFT nuisance parameters~\cite{Chudaykin:2020hbf,Chudaykin:2020ghx}, 
\be
\{\omega_{cdm},H_0,A_s,n_s\}\times \{b_1,b_2,b_{\mathcal{G}_2},b_{\Gamma_3},c_0,c_2,b_4,a_0,a_2,P_{\rm shot},B_{\rm shot},c_1\}\,.
\ee
The priors on the power spectrum nuisance parameters 
are given in Appendix~\ref{app:power}. As for $B_{\rm shot}$,
we place a Gaussian prior on it with unit mean, which corresponds 
to the Poissonian sampling prediction,
and unit variance,
\be 
B_{\rm shot}\sim \mathcal{N}(1,1^2)\,.
\ee
$c_1$ is varied in our MCMC chains without any priors, unless
otherwise stated.
We fix the physical baryon density to its 
true value in order to simulate the BBN prior
as it was used in Refs.~\cite{Ivanov:2019pdj,Chudaykin:2020ghx}.\footnote{Formally, we also use the FIRAS value of the current CMB temperature $T_0$,
which is a required input parameter 
in the Boltzmann code \texttt{CLASS}~\cite{Blas:2011rf}.
This parameter is tightly constrained 
by FIRAS and other probes, see e.g.~\cite{Ivanov:2020mfr} for more detail.
}

\section{Results}
\label{sec:results}

We start our analysis from the simple case of the real space bispectrum, 
which is free from RSD and projection effects. 
Then, we will analyze a setup 
that closely matches an actual spectroscopic survey:
we will study the bispectrum in redshift space and in 
the presence of the projection effect.
Since the PT challenge data which we are using is still ongoing, we report the measurements of all cosmological 
parameters normalized to their 
true injected values.
As far as nonlinear bias parameters are concerned, 
we will present their values after the subtraction of
the fiducial values extracted from our best-fit estimates 
from the most constraining baseline likelihood analysis. 
Specifically, we will report 
\be 
\Delta b_2 \equiv b_2 - b^{\rm bf}_2\,,\quad 
\Delta b_{\mathcal{G}_2} \equiv b_{\mathcal{G}_2} - b^{\rm bf}_{\mathcal{G}_2}\,,\
\ee
where $b^{\rm bf}_2, b^{\rm bf}_{\mathcal{G}_2}$ are best-fit values 
extracted from the fiducial analysis of the redshift space 
power spectrum combined with the
real space bispectrum at $k_{\rm max}=0.08~h$/Mpc.
This will be our best guess for the true values of 
these parameters.

We emphasize that expect for Section~\ref{sec:boss}, 
in all our analysis the 
scale cuts of the power spectrum likelihood are kept fixed.
Only $\kmax$ of the bispectrum data is varied. 

\begin{figure}
\centering
\includegraphics[width=0.99\textwidth]{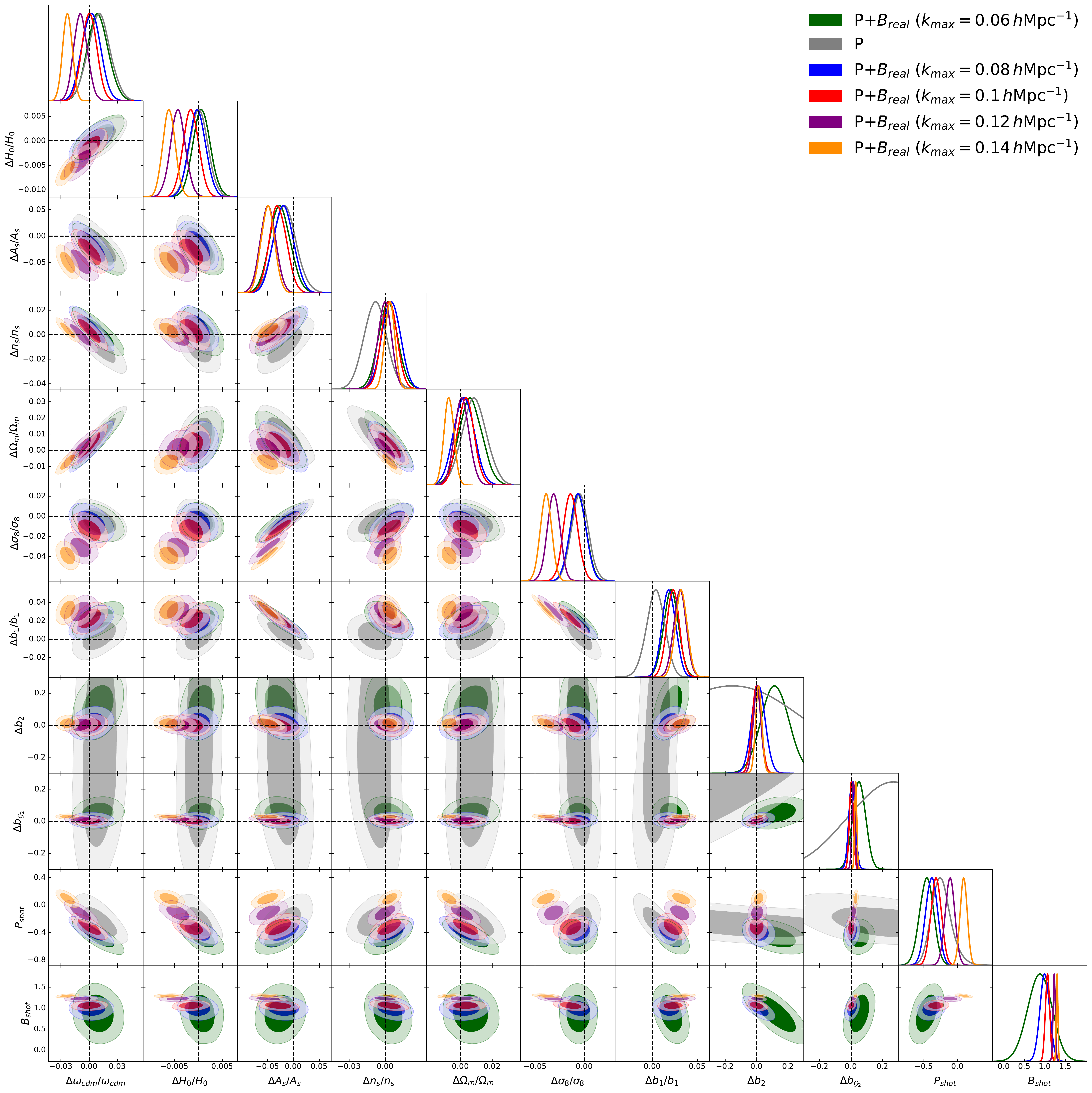}
\caption{Posterior distributions of cosmological 
and some nuisance parameters from MCMC analyses of 
the joint redshift-space power spectrum and real space 
bispectrum data. We show results for five different choices 
of the bispectrum data cut $\kmax$. All cosmological parameters 
and $b_1$ are normalized to their true values.
We have subtracted constant fiducial values from the quadratic 
bias parameters $b_2$ and $b_{\mathcal{G}_2}$. Results for the power
spectrum data only are shown for comparison.
}
\label{fig:real}
\end{figure}

\begin{figure}[ht!]
\centering
\includegraphics[width=0.49\textwidth]{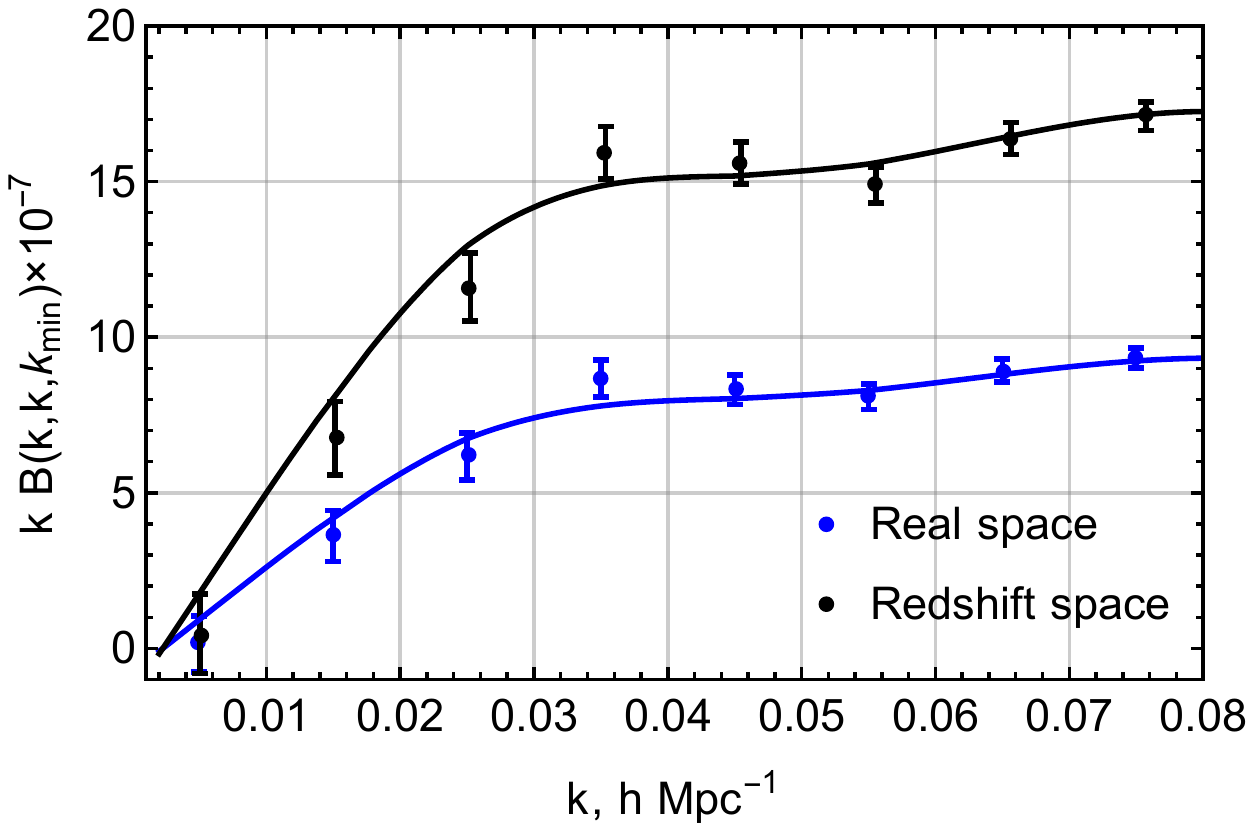}
\includegraphics[width=0.49\textwidth]{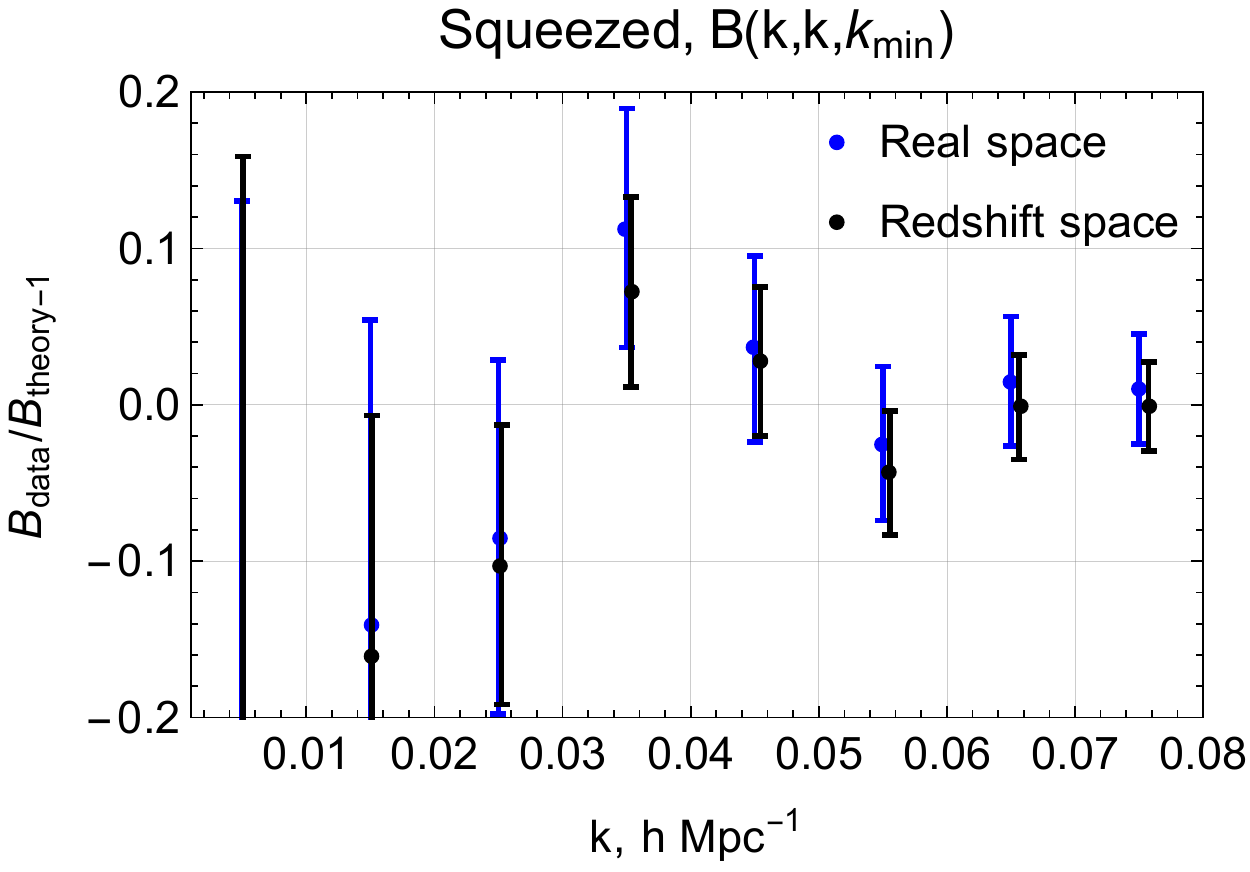}
\includegraphics[width=0.49\textwidth]{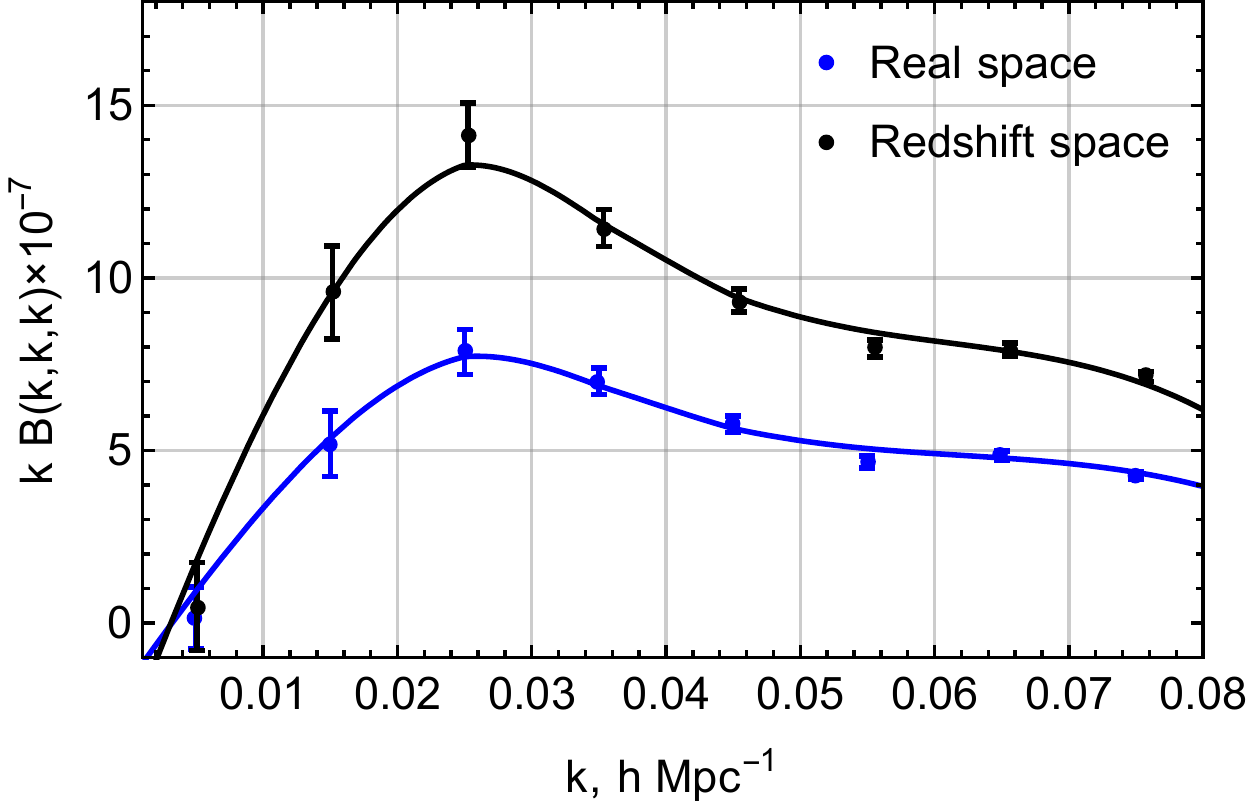}
\includegraphics[width=0.49\textwidth]{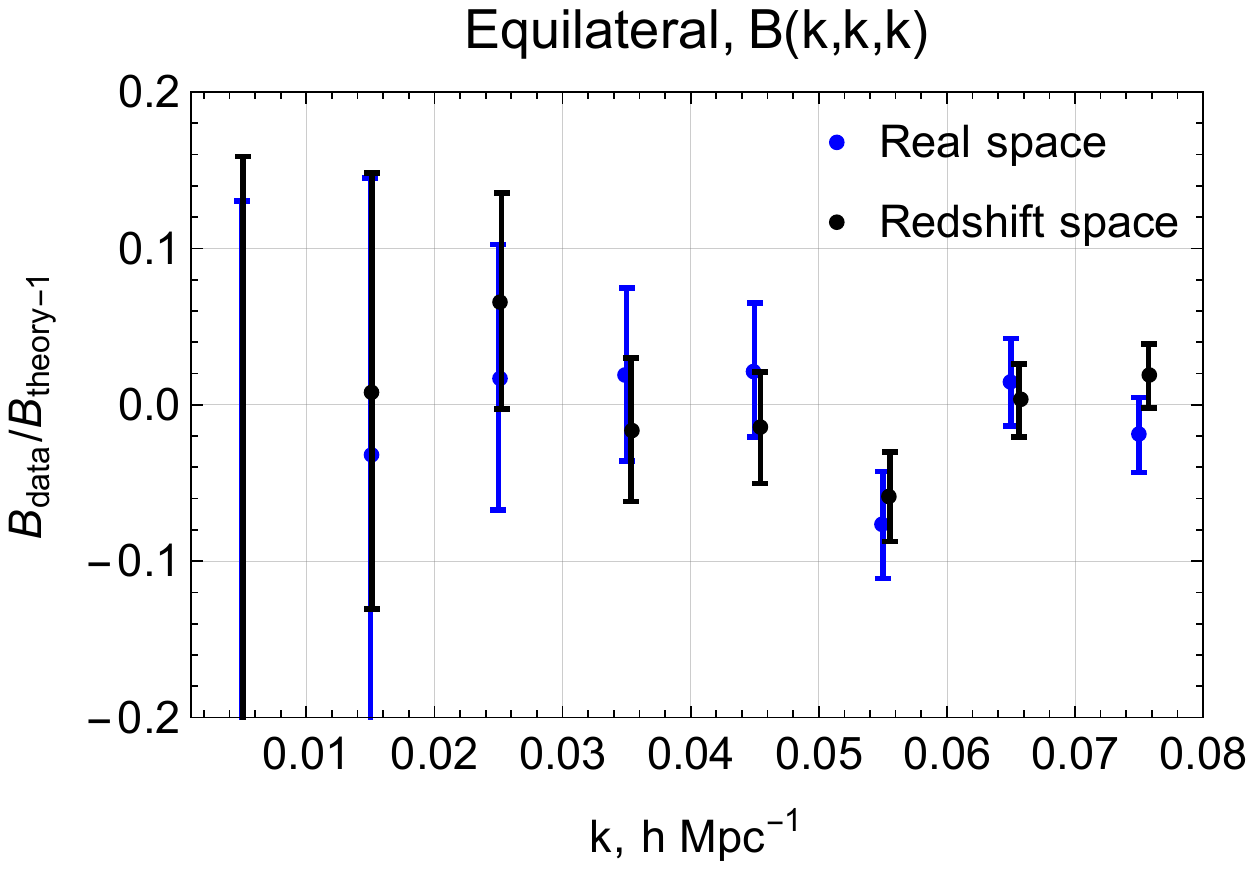}
\includegraphics[width=0.49\textwidth]{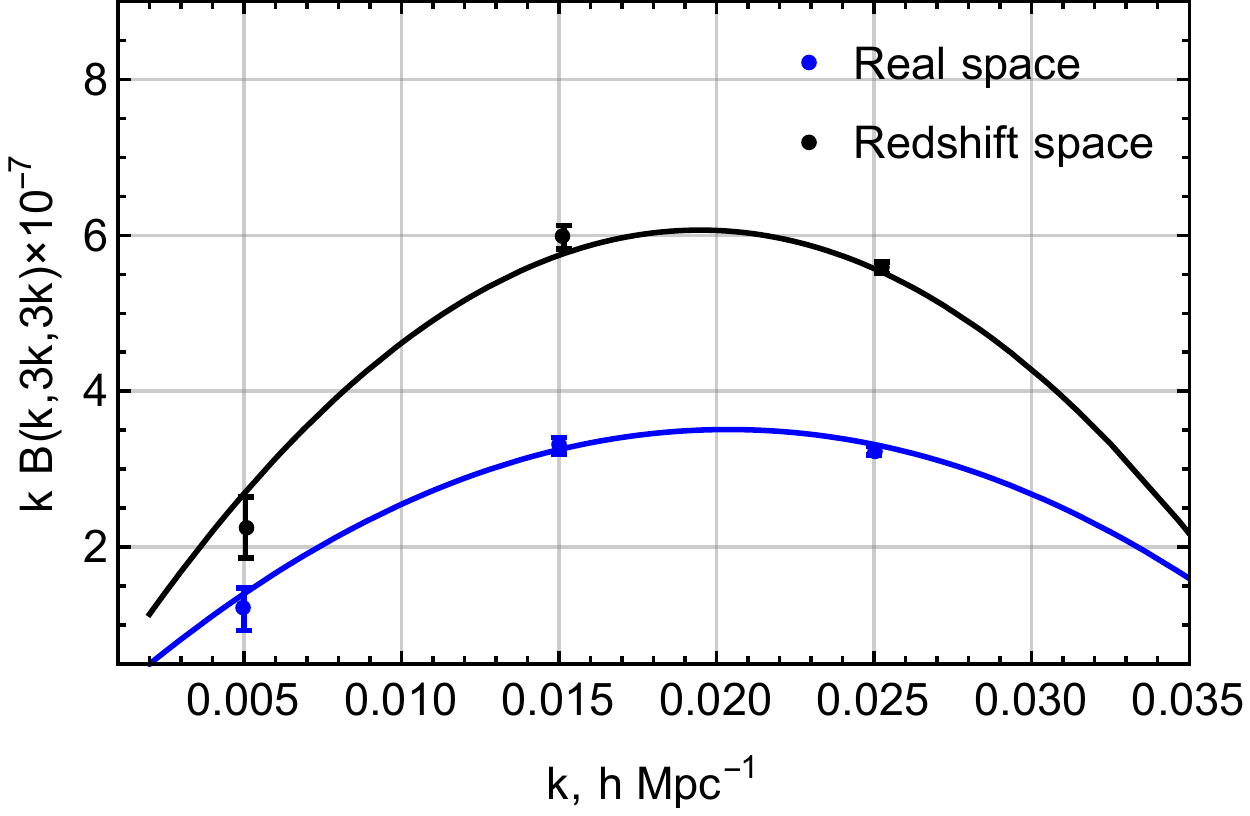}
\includegraphics[width=0.49\textwidth]{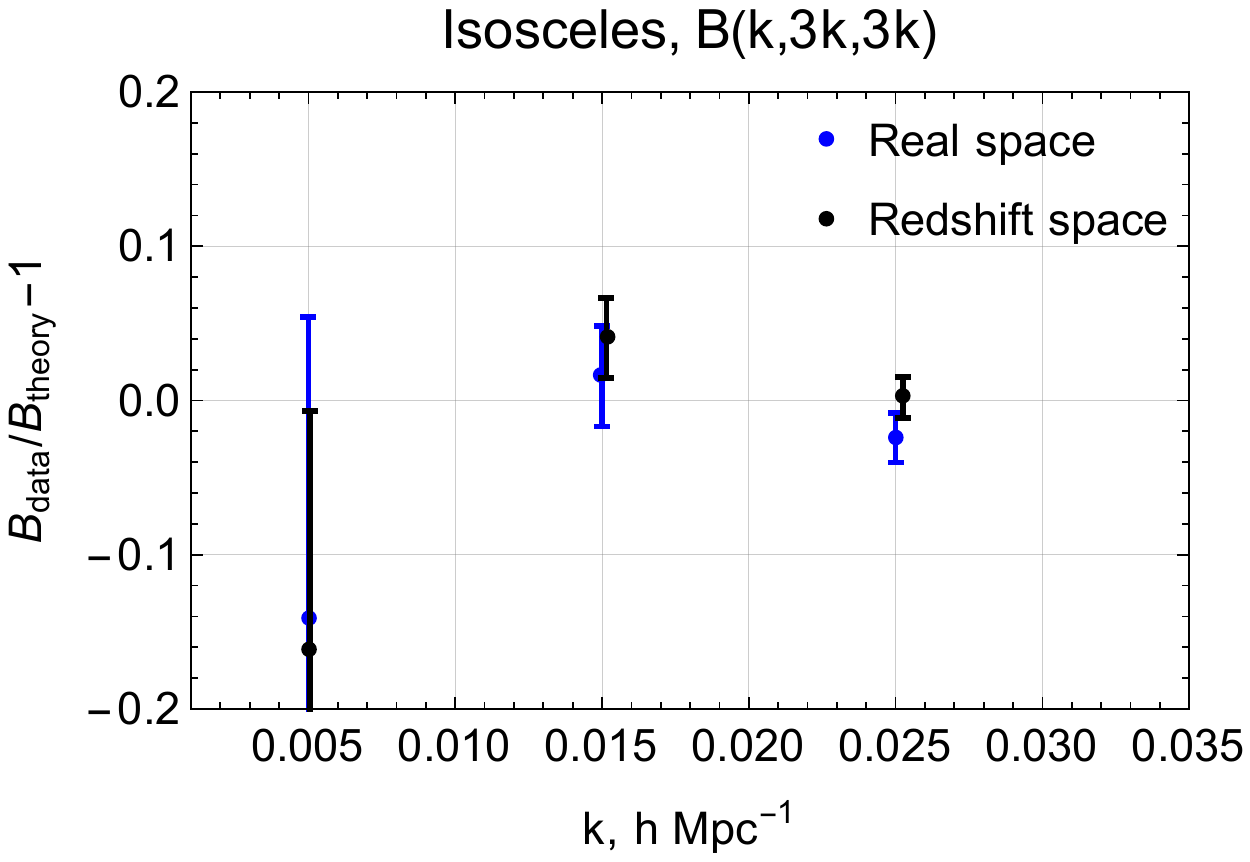}
\caption{
Bispectrum data points from the PT challenge simulations along with 
best-fit theoretical predictions extracted from our MCMC chains.
We show the bispectra for squeezed, equilateral and isosceles triangles 
(left panels), and the corresponding residuals (right panels).
}
\label{fig:data}
\end{figure}

\begin{figure}[ht!]
\centering
\includegraphics[width=0.49\textwidth]{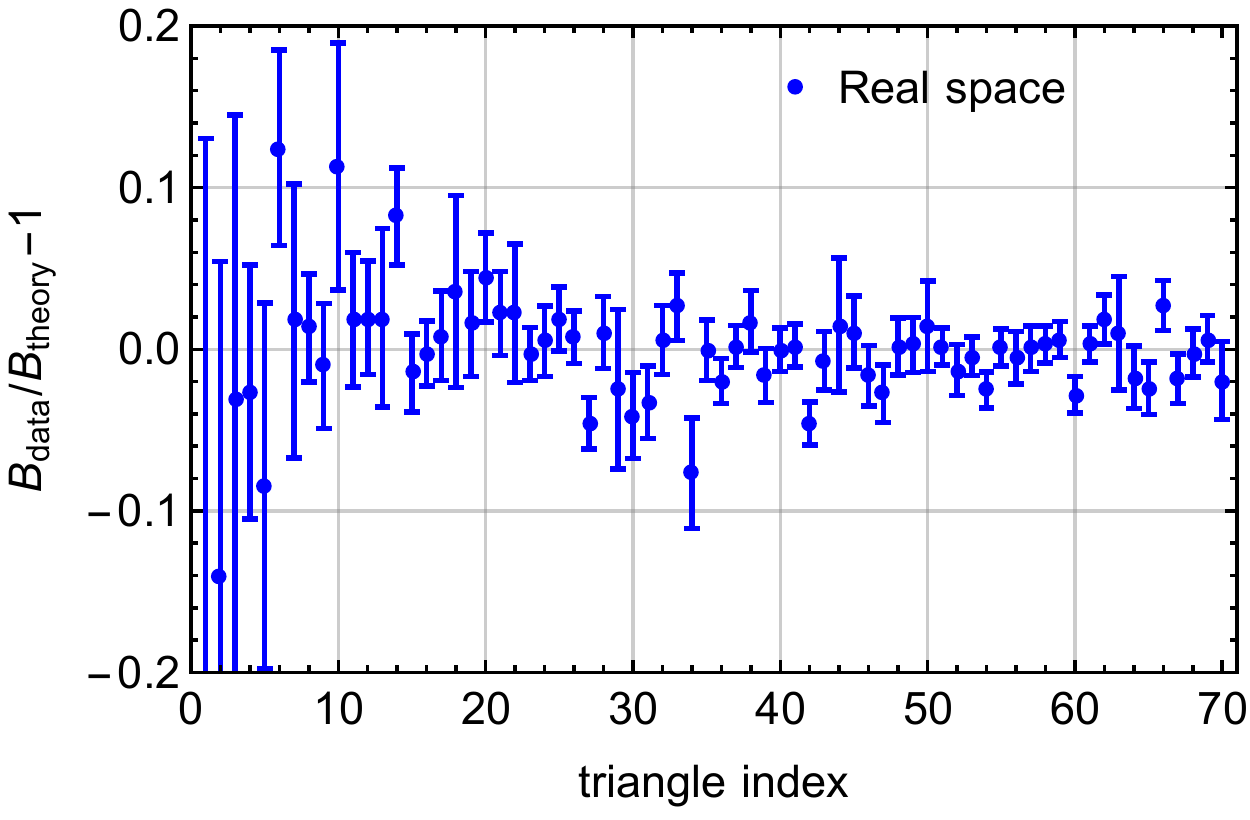}
\includegraphics[width=0.49\textwidth]{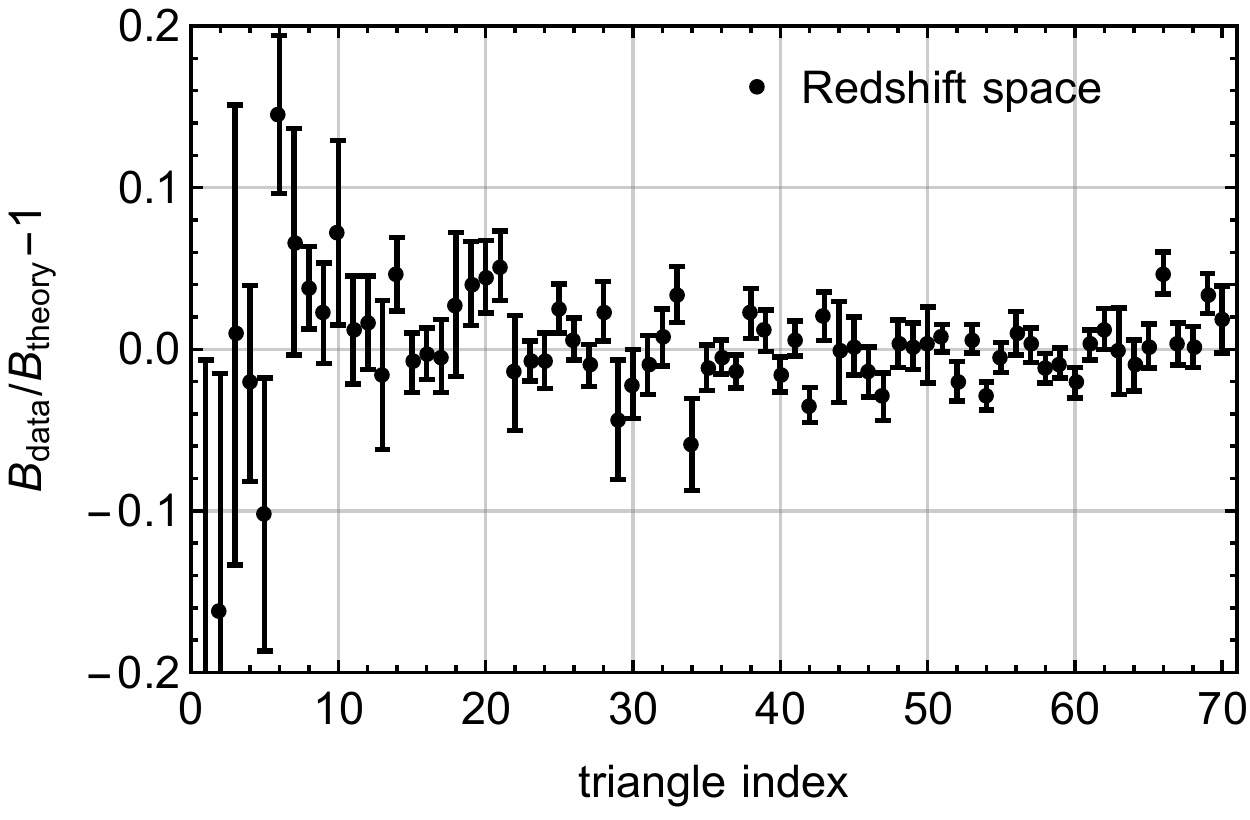}
\caption{
Residuals between the bispectrum data 
and best-fitting theory templates 
for all triangles from the real space (left panel),
and the redshift space (right panel) analyses.
}
\label{fig:resid}
\end{figure}

\subsection{Redshift space power spectrum + real space bispectrum}

We start with the real space bispectrum, which can be formally 
obtained from our model Eq.~\eqref{eq:Bfull}
by setting $f=0,~c_1=0$, and ignoring the AP effect. 
We also note that the discreteness weights 
are closer to unity in this case. This can be attributed 
to the absence of leakage from higher angular moments~\cite{Tegmark:2003uf},
which is present in the redshift-space case. 
We perform our analysis for the bispectrum for five choices of $\kmax$  
ranging from $0.06$ to $0.14~\hMpc$ with a step $0.02~\hMpc$.
The resulting corner plot from our MCMC analyses is shown in Fig.~\ref{fig:real}, and the 1d marginalized 
limit for the case $\kmax=0.08\hMpc$ are presented in Table~\ref{tab:final}. 
The best fitting curves for certain triangle configurations are shown in Fig.~\ref{fig:data}, while Fig.~\ref{fig:resid} is a residual plot over all triangles used in the fit.

\begin{table}[!htb]
    % \caption{Global caption}
    \begin{minipage}{0.5\linewidth}
\vspace{0.1cm}
\begin{tabular}{|l|c|c|}
\hline
\multicolumn{2}{|c|}{\text{Power}}
\\ \hline
Parameter & 68\% limits \\ \hline
\hline

$\Delta H_0/H_0            $ & $0.0001\pm 0.0019          $\\

$\Delta \omega_{cdm}/\omega_{cdm}$ & $0.010\pm 0.012            $\\

$\Delta A_s/A_s            $ & $-0.016\pm 0.022           $\\

$\Delta n_s/n_s            $ & $-0.0084\pm 0.0094         $\\

$\Delta b_1/b_1            $ & $0.003\pm 0.010            $\\

$\Delta b_2                $ & $-0.13^{+0.44}_{-0.51}     $\\

$\Delta b_{\mathcal{G}_2}  $ & $0.29\pm 0.30              $\\

{$P_{{shot} }    $} & $-0.24^{+0.12}_{-0.16}     $\\
\hline 
$\Delta \sigma_8/\sigma_8  $ & $-0.0045\pm 0.0087         $\\

$\Delta \Omega_m/\Omega_m  $ & $0.0087\pm 0.0077          $\\

 $\frac{b_1^3 \sigma_8^4}{(b_1^3 \sigma_8^4)_{\rm fid.}} -1$
 & $-0.008\pm 0.022 $ \\

\hline

\end{tabular}
 \end{minipage}%b
\begin{minipage}{.5\linewidth}
\begin{tabular}{|l|c|c|}
\hline
\multicolumn{2}{|c|}{\text{Power+real space bispectrum}}
\\ \hline
Parameter & 68\% limits \\ \hline
\hline

$\Delta H_0/H_0            $ & $-0.0002\pm 0.0018         $\\

$\Delta \omega_{cdm}/\omega_{cdm}$ & $0.0022\pm 0.0098          $\\

$\Delta A_s/A_s            $ & $-0.019\pm 0.019           $\\

$\Delta n_s/n_s            $ & $0.0048\pm 0.0076          $\\

$\Delta b_1/b_1            $ & $0.0183\pm 0.0077          $\\

$\Delta b_2                $ & $0.011\pm 0.043            $\\

$\Delta b_{\mathcal{G}_2}  $ & $0.006\pm 0.020            $\\

{$P_{{shot} }    $} & $-0.384\pm 0.089           $\\

{$B_{{shot} }    $} & $0.99\pm 0.12              $\\

\hline

$\Delta \sigma_8/\sigma_8  $ & $-0.0065\pm 0.0080         $\\

$\Delta \Omega_m/\Omega_m  $ & $0.0022\pm 0.0064          $\\

 $\frac{b_1^3 \sigma_8^4}{(b_1^3 \sigma_8^4)_{\rm fid.}} -1$
 & $0.028\pm 0.016$ \\

\hline
\end{tabular}
 \end{minipage}%b
 \\
    \begin{minipage}{.5\linewidth}
     % \vspace{-0.1cm}
\begin{tabular}{|l|c|c|}
\hline
\multicolumn{2}{|c|}{\text{RSD bispectrum}}
\\ \hline
Parameter & 68\% limits \\ \hline
\hline
$\Delta H_0/H_0            $ & $-0.026\pm 0.015           $\\

$\Delta \omega_{cdm}/\omega_{cdm}$ & $-0.026\pm 0.032           $\\

$\Delta A_s/A_s            $ & $-0.07^{+0.17}_{-0.35}     $\\

$\Delta n_s/n_s            $ & $-0.018\pm 0.034           $\\

$\Delta b_1/b_1            $ & $0.07^{+0.20}_{-0.26}      $\\

$\Delta b_2                $ & $0.81^{+0.26}_{-0.33}      $\\

$\Delta b_{\mathcal{G}_2}  $ & $0.270^{+0.068}_{-0.091}   $\\

$c_1                       $ & $-2.4\pm 3.8               $\\

{$P_{{shot} }    $} & $-0.095\pm 0.93            $\\

{$B_{{shot} }    $} & $0.89^{+0.69}_{-0.61}      $\\

\hline

$\Delta \sigma_8/\sigma_8  $ & $-0.07^{+0.11}_{-0.16}     $\\

$\Delta \Omega_m/\Omega_m  $ & $0.033^{+0.032}_{-0.039}   $\\

 $\frac{b_1^3 \sigma_8^4}{(b_1^3 \sigma_8^4)_{\rm fid.}} -1$
 & $-0.194^{+0.082}_{-0.074}$ \\

\hline
\end{tabular}
\end{minipage}%
 % \centering{
    \begin{minipage}{.5\linewidth}
     % \vspace{-0.1cm}
\begin{tabular}{|l|c|c|}
\hline
\multicolumn{2}{|c|}{\text{Power+RSD bispectrum}}
\\ \hline
Parameter & 68\% limits \\ \hline
\hline

$\Delta H_0/H_0            $ & $-0.0014\pm 0.0018         $\\

$\Delta \omega_{cdm}/\omega_{cdm}$ & $-0.005\pm 0.010           $\\

$\Delta A_s/A_s            $ & $-0.017\pm 0.021           $\\

$\Delta n_s/n_s            $ & $0.0036\pm 0.0080          $\\

$\Delta b_1/b_1            $ & $0.0149\pm 0.0085          $\\

$\Delta b_2                $ & $-0.054\pm 0.088           $\\

$\Delta b_{\mathcal{G}_2}  $ & $0.070\pm 0.026            $\\

$c_1                       $ & $5.6\pm 2.7                $\\

{$P_{{shot} }    $} & $-0.249\pm 0.093           $\\

{$B_{{shot} }    $} & $1.75\pm 0.43              $\\
\hline
$\Delta \sigma_8/\sigma_8  $ & $-0.0107\pm 0.0082         $\\

$\Delta \Omega_m/\Omega_m  $ & $-0.0017\pm 0.0068         $\\

 $\frac{b_1^3 \sigma_8^4}{(b_1^3 \sigma_8^4)_{\rm fid.}} -1$
 & $0.001\pm 0.016$ \\

\hline
\end{tabular}
\end{minipage}%
% }
\caption{1d marginalized limit for the cosmological and most important nuisance 
parameters from various PT challenge likelihoods: 
redshift space power spectrum only
(upper left panel), 
the joint power spectrum + real space bispectrum (upper right panel), 
redshift space bispectrum only (lower left panel),
and the joint redshift space 
power spectrum + redshift space bispectrum (lower right panel).
Parameters that were directly varied in MCMC chains are displayed 
in the upper part of the table, 
the lower groups contain derived parameters. Most parameters are normalized to their true 
values. See the main text for more detail.
\label{tab:final}
}
\end{table}

We observe that 
inclusion of the bispectrum
sharpens estimates for all cosmological 
and bias parameters 
and does not lead to any significant biases up to $\kmax=0.08\hMpc$.
We see some small biases, especially in the $b_1-\sigma_8$ plane, 
but our MCMC posteriors still enclose the true values within $99\%$ CL,
which makes these shifts compatible with statistical fluctuations.
Besides, these small shifts do not change when switching 
the bispectrum data cut from $0.06\hMpc$ to $0.08\hMpc$.
In contrast, for $\kmax>0.08\hMpc$ we see clear shifts that push estimated 
values away from the ground truth. In particular, we find
the bias on $\sigma_8$ to be $[-1.9,-5.2,-6.7]\sigma$
for $\kmax/\hMpc=[0.1,0.12,0.14]$, respectively.
This suggests us to adopt 
$\kmax=0.08\hMpc$ 
as a baseline data cut for the real space bispectrum 
in what follows. 

The tree-level bispectrum likelihood improves 
constraints on cosmological and some nuisance parameters. 
This improvement can be estimated by ratios of the 
1d marginalized $68\%$ confidence intervals. 
For the cosmological parameters 
we have
\[
\frac{\sigma_{\rm P+B}}{\sigma_{\rm P}}\{\omega_{cdm},h,n_s,A_s,\Omega_m,\sigma_8\}=\{0.82,0.90,0.81,0.88,0.83,0.93\}\,
\]
indicating a $(10-20)\%$ improvement in most cases. 
The gain is more sizable for the nuisance parameters,

\[
\frac{\sigma_{\rm P+B}}{\sigma_{\rm P}}
\{b_1,b_2,b_{\mathcal{G}_2},P_{\rm shot}\}
=\{0.75,0.09,0.07,0.61\}\,.
\]
Intuitively, this happens because in the bispectrum one can probe
the galaxy bias parameters
from large scales,
and hence their 
determination
is not contaminated
by loop corrections
and additional
nuisance parameter marginalization. 

The picture that we have observed here is in stark
contrast with the real space only results, 
presented in Appendix~\ref{app:real}. 
This analysis shows that the 
real space power spectrum has much less 
information than the redshift-space one.
In this case the combination
with the bispectrum leads to a dramatic 
shrinking of posterior distributions for 
both cosmological and nuisance parameters.
However, in redshift space the power spectrum
has much more information to begin with,
and thus the addition of the bispectrum 
yields only 
a moderate improvement.

\subsection{Bias parameters}

Our simulated 
galaxies
are produced 
with 
simple HOD models
and therefore one may expect their nonlinear bias parameters to match those of the host halos
and to follow the same dependence on $b_1$. 
Let us compare this expectations with reality.
For the tidal bias $b_{\mathcal{G}_2}$, as a first guess, we can use the 
so-called Lagrangian Local In Matter Density (LLIMD) bias model prediction 
$b_{\mathcal{G}_2}^{\rm LLIMD}=-2(b_1-1)/7$~\cite{Desjacques:2016bnm}.
Using the fiducial value of $b_1$ we find 
\be 
\Delta b_{\mathcal{G}_2}^{\rm LLIMD}=-\frac{2}{7}(b^{\rm fid}_1-1)-
b_{\mathcal{G}_2}^{\rm fid}=0.23\,,
\ee
which is more that $10\sigma$ away from the truth. The LLIMD approximation is known to be in tension with high precision 
simulation measurements, which clearly show the evidence for 
the tidal Lagrangian bias~\cite{Abidi:2018eyd}. A better fit to this data 
is a coevolution model with the initial Lagrangian bias 
that has the following dependence on the mean halo mass $M$
\be
b^L_{\mathcal{G}_2} = -0.5\left(\frac{M}{4\times 10^{14}h^{-1}M_{\odot}}\right)^{0.8}\,.
\ee
 Ref.~\cite{Abidi:2018eyd} also presents 
the function $M(b_1)$, from which we can express the above equation
as $b^L_{\mathcal{G}_2}(b_1)$.
Inserting there our measurement of $b_1$, we find
\be 
\Delta b_{\mathcal{G}_2}^{\rm LTCM}=
-\frac{2}{7}(b^{\rm fid}_1-1)
+b^L_{\mathcal{G}_2}(b^{\rm fid}_1)
-
b_{\mathcal{G}_2}^{\rm fid}=0.072\,,
\ee
where `LTCM' stands for `Lagrangian tidal coevolution model'. 
We see that our measurement is still in $\sim 3\sigma$
tension with the prediction of LTCM, although in absolute terms 
the discrepancy is quite small. The discrepancy 
with the excursion set prediction from 
Ref.~\cite{Eggemeier:2020umu} is 
also quite high, it exceeds 10$\sigma$ in terms of 
the standard deviation of our measurement.

As far as $b_2$ is concerned, we can compare our measurement 
with the fit to halos from Refs.~\cite{Lazeyras:2015lgp,Lazeyras:2017hxw}, i.e.~to consider
\be 
\Delta b_{2}^{\rm halo}=b^{\rm halo}_2(b^{\rm fid}_1,b^{\rm fid}_{\mathcal{G}_2})-
b_{2}^{\rm fid}=-0.49\,,
\ee
where 
\be 
b^{\rm halo}_2(b_1,b_{\mathcal{G}_2})= 
0.412 - 2.143 b_1 +0.929 b_1^2 +0.008 b_1^3
+\frac{4}{3}b_{\mathcal{G}_2}\,.
\ee
Note that we have accounted for the difference in our definition of 
quadratic biases w.r.t.~Refs.~\cite{Lazeyras:2015lgp,Lazeyras:2017hxw},
\be
b^{\rm this\,work}_{\mathcal{G}_2}=b^{\rm previous}_{\mathcal{G}_2}\,,
\quad b_2^{\rm this\,work}= b_2^{\rm previous}+\frac{4}{3}b^{\rm this\,work}_{\mathcal{G}_2}\,.
\ee
Thus, our analysis confirms significant 
deviations between the bias coefficients of galaxies 
and halos, which have already been pointed out
in the literature~\cite{Eggemeier:2021cam,Barreira:2021ukk}. We also confirm the trend seen 
in the literature for the CMASS-like galaxies~\cite{Alam:2016hwk}(similar to our PT challenge sample): the tidal bias of galaxies is lower 
than that of halos, but $b_2$ is higher. 
In particular, the results of Ref.~\cite{Eggemeier:2021cam} 
for the CMASS galaxies
read
\be
b_2^{\rm gal}=-0.2\pm 0.1\,,
\quad  
b^{\rm gal}_{\mathcal{G}_2}=-0.46\pm 0.06\,.
\ee
These values can be compared with 
the predictions of the local Lagrangian 
approximation and the fit to $b_2$,
\be 
\begin{split}
\text{Ref.~\cite{Eggemeier:2021cam}:}\quad & \Delta b_{2}=b^{\rm halo}_{2}(b^{\rm gal}_{2},b^{\rm gal}_{\mathcal{G}_2})-b^{\rm gal}_{2}= -0.41 \pm 0.1\,,\\
& \Delta b_{\mathcal{G}_2}=b^{\rm LLIMD}_{\mathcal{G}_2}
-b^{\rm gal}_{\mathcal{G}_2}= 0.17\pm 0.06 \,.
\end{split}
\ee
These estimates perfectly agree with our results 
\be \Delta b_2=-0.49\pm 0.04\,, \quad 
\Delta b_{\mathcal{G}_2}=0.23\pm 0.02\,. \ee

Finally, let us discuss the cubic tidal bias parameter
$b_{\Gamma_3}$. At the power spectrum level
it is almost fully degenerate with $b_{\mathcal{G}_2}$.
However, this degeneracy gets broken by the bispectrum 
data, since only $b_{\mathcal{G}_2}$ enters the tree-level bispectrum model. 
We will compare our measurements with 
halo relations obtained in Refs.~\cite{Abidi:2018eyd,Eggemeier:2020umu},
\be
\begin{split}
\text{Ref.~\cite{Abidi:2018eyd}:} \quad & b_{\Gamma_3}^{\rm halo}=
-b_{\mathcal{G}_2}-\frac{1}{15}(b_1-1) \,,\\
\text{Ref.~\cite{Eggemeier:2020umu}:} \quad & 
b_{\Gamma_3}^{\rm halo}=
-\frac{1}{6}(b_1-1)-\frac{3}{2}b_{\mathcal{G}_2}\,.
\end{split}
\ee
This gives us a tension between our results and these halo predictions at the 2$\sigma$
level, 
\be
\begin{split}
\text{Ref.~\cite{Abidi:2018eyd}:} \quad & 
b_{\Gamma_3}^{\rm gal}-b_{\Gamma_3}^{\rm halo}(b_1^{\rm gal},b_{\mathcal{G}_2}^{\rm gal})=0.23 \pm 0.11
 \,,\\
\text{Ref.~\cite{Eggemeier:2020umu}:} \quad & 
b_{\Gamma_3}^{\rm gal}-b_{\Gamma_3}^{\rm halo}(b_1^{\rm gal},b_{\mathcal{G}_2}^{\rm gal})=
0.24 \pm 0.11\,.
\end{split}
\ee
However, here we see the difference w.r.t.~the CMASS-like sample
of Ref.~\cite{Eggemeier:2021cam} 
($b_{\Gamma_3}=-7\gamma_{21}/4$
in their notation).
The relevant measurement from 
this work is fully consistent with that of halos,
\be 
b^{\rm CMASS}_{\Gamma_3} - b_{\Gamma_3}^{\rm halo}
(b_1^{\rm gal},b_{\mathcal{G}_2}^{\rm gal})=-0.02 \pm 0.14\,.
\ee
The discrepancy between our $b_{\Gamma_3}$
and that of Ref.~\cite{Eggemeier:2021cam} 
is marginally below $2\sigma$, and hence our measurements
can be considered consistent.

Overall, we conclude that with the PT challenge simulations
we see a $\sim 3\sigma$ discrepancy between 
the bias parameters of our HOD galaxies and their host halos.
On the one hand, our bias parameter measurements
agree well 
with those from similar mock CMASS-like galaxies,
analyzed 
in Ref.~\cite{Eggemeier:2021cam}.

\begin{figure}
\centering
\includegraphics[width=0.99\textwidth]{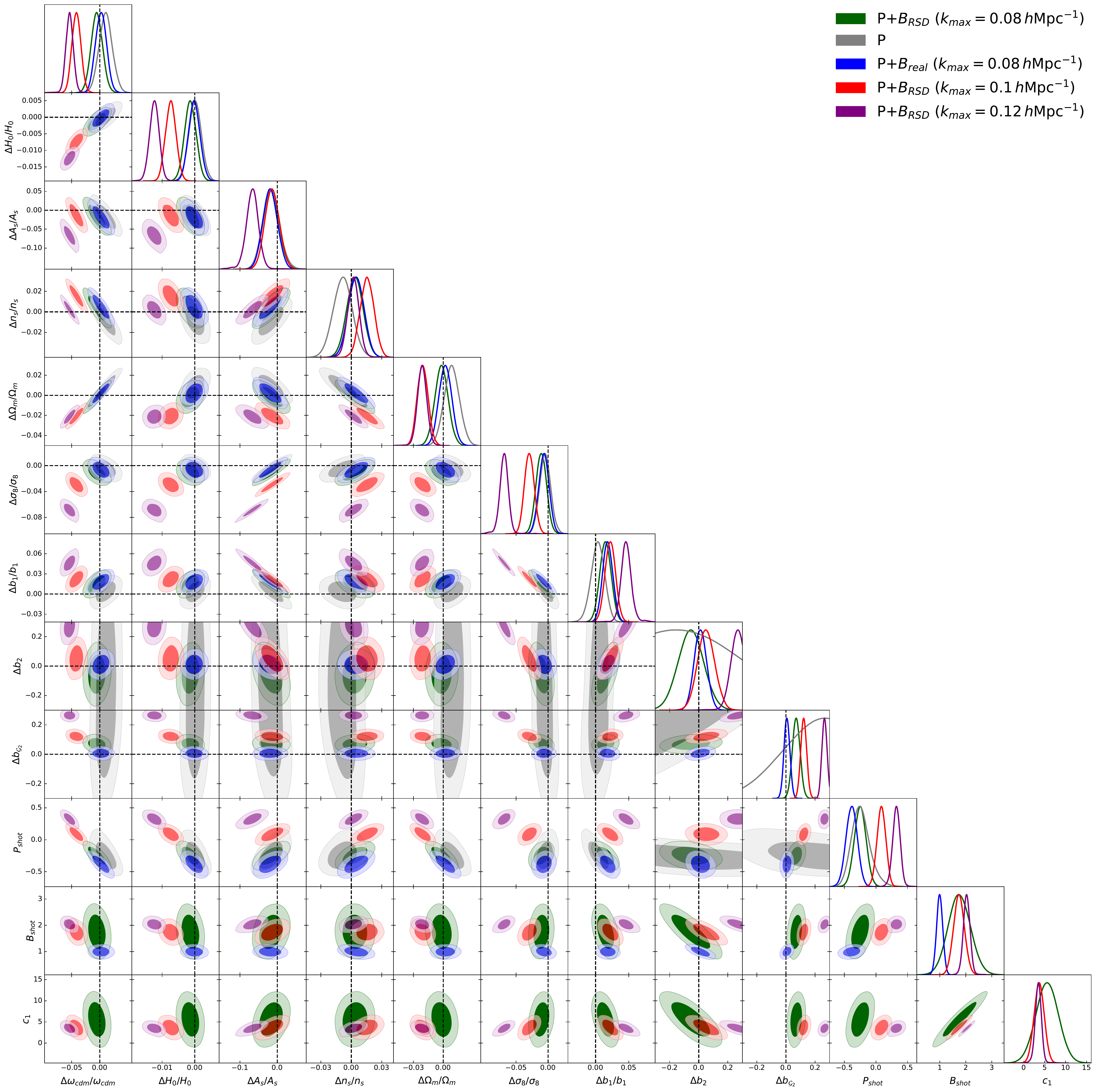}
\caption{
Posterior distributions of cosmological 
and certain nuisance parameters from MCMC analyses of 
the joint redshift space power spectrum and redshift space 
bispectrum monopole data. We show results for three different choices 
of the bispectrum data cut $\kmax$. All cosmological parameters 
and $b_1$ are normalized to their true values.
We have subtracted constant fiducial values from the quadratic 
bias parameters $b_2$ and $b_{\mathcal{G}_2}$. Results for the power
spectrum (``$P$'') and for the power spectrum+real space bispectrum (``$P+B_{\rm real}~(\kmax=0.08~\hMpc)$'')
datasets
are shown for comparison.
}
\label{fig:rsd}
\end{figure}

\begin{figure}
\centering
\includegraphics[width=0.99\textwidth]{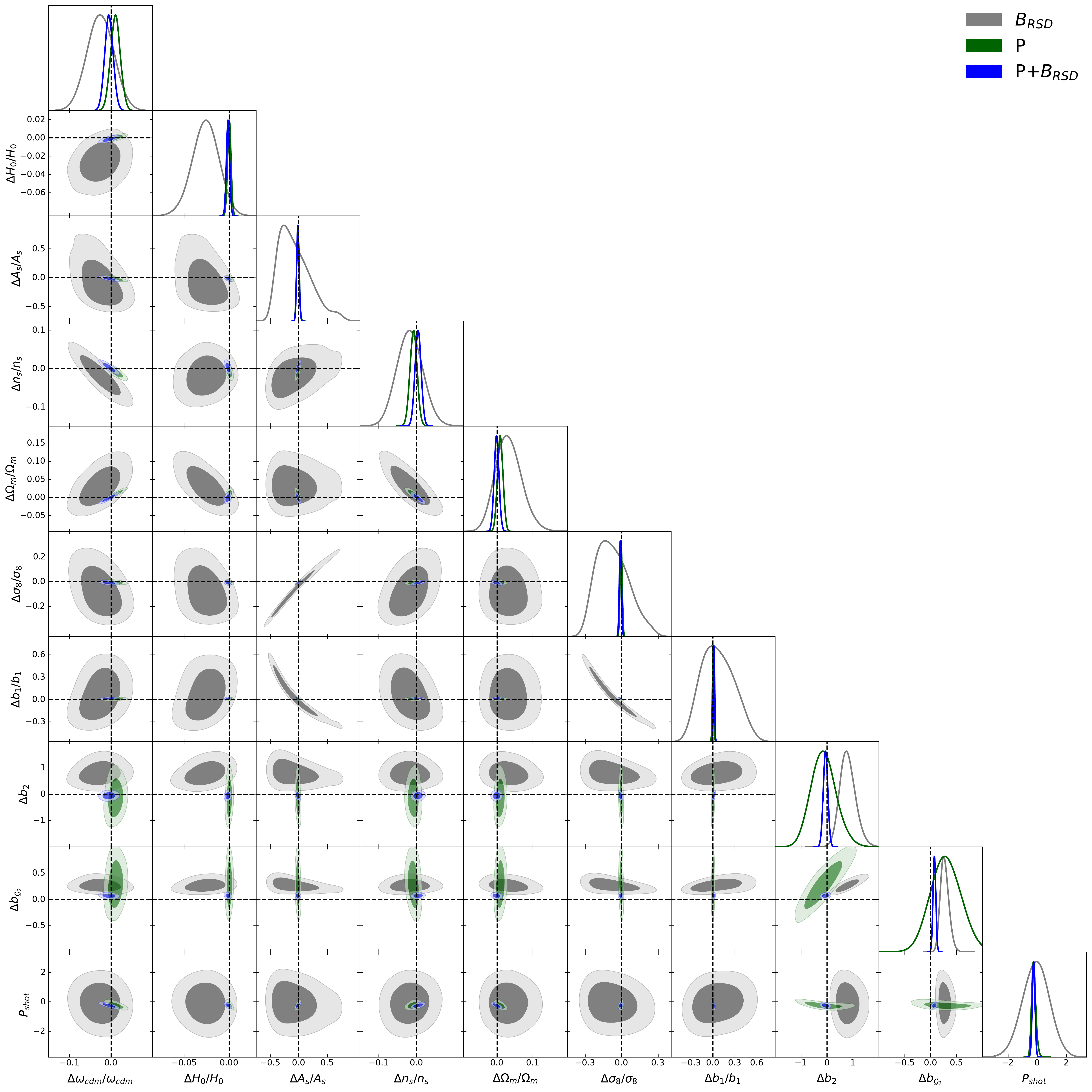}
\caption{
Posterior distributions of cosmological 
and certain nuisance parameters from MCMC analyses of 
the redshift space power spectrum, redshift space bispectrum
and their combination. We use $\kmax=0.08~\hMpc$
for the bispectrum here.
}
\label{fig:rsdONLY}
\end{figure}

\subsection{Redshift space}

We now consider the realistic case of the redshift-space bispectrum
monopole in the presence of the AP effect. We analyze our joint 
power spectrum and bispectrum likelihoods for three choices 
of the bispectrum data cut ranging
$\kmax$ from $0.08~\hMpc$ to $0.12~\hMpc$ with a step $0.02~\hMpc$.
Our triangle plot is displayed in Fig.~\ref{fig:rsd}, where for comparison 
we also show the results of the baseline real space bispectrum
analysis from the previous section. Marginalized 1d limits are presented in
Table~\ref{tab:final}. Best-fit curves and the residual plot are shown in Fig.~\ref{fig:data}
and in Fig.~\ref{fig:resid}.

We see that at $\kmax=0.08~\hMpc$ the 
addition of the bispectrum likelihood  
slightly
narrows the power spectrum contours and does not lead to any significant bias. 
Both cosmological and nuisance parameters are recovered 
within $95\%$ confidence intervals. However, already at 
$\kmax=0.08~\hMpc$ we observe some evidence for the non-zero 
FoG counterterm $c_1$, which suggests 
that the one-loop corrections may not be negligible.
Indeed, for more aggressive data cuts $\kmax>0.08~\hMpc$
we find large biases that signal the breakdown 
of the tree-level bispectrum model.
These biases are more significant than those 
that we have seen in the real space power spectrum likelihood,
which is an expected consequence of non-linear
redshift space distortions~\cite{Scoccimarro:1999ed,Pueblas:2008uv,Lewandowski:2015ziq,Chudaykin:2020hbf,Schmittfull:2020trd,Chen:2020zjt}. 
A similar conclusion that
FoGs in the bispectrum are important even on relatively large scales 
was made in Ref.~\cite{Smith:2007sb}.
These results motivate us to choose $\kmax=0.08~\hMpc$ as our baseline data cut.

In Fig.~\ref{fig:rsdONLY} and table~\ref{tab:final}
we display the breakdown of different 
likelihoods in terms of their parameter constraints, 
including the redshift space bispectrum alone.
Clearly, the constraints on cosmological parameters 
are heavily 
dominated by the power spectrum data.
In part, this is a result of using only 
relatively low
wavenumbers in our bispectrum analysis.

The bispectrum likelihood adds
new information mostly through the bias
parameter measurements. 
In particular, the principle component of the 
$b_1-\sigma_8$ degeneracy can be well approximated by 
a combination 
$b_1^3 \sigma_8^4$, which captures the 
galaxy bispectrum amplitude 
in the absence of quadratic biases
and projection effects. 
Our redshift space bispectrum-only analysis yields
a measurement quite competitive 
with the redshift-space power
spectrum result\footnote{For the power spectrum the principle
component is slightly different, $b_1^2 \sigma_8^4$. This small difference in the exponent is not important for our discussion.},
c.f. table~\ref{tab:final}.
Beside $b_1-\sigma_8$,
the bispectrum also adds significant information 
through the quadratic 
bias parameters $b_2$ and $b_{\mathcal{G}_2}$,
whose measurements from the bispectrum alone
are more precise than from the power spectrum.

Addition of the bispectrum leads to following improvements on cosmological and nuisance parameters 
\be 
\begin{split}
&\frac{\sigma_{\rm P+B}}{\sigma_{\rm P}}\{\omega_{cdm},h,n_s,A_s,\Omega_m,\sigma_8\}=\{0.88,0.94,0.86,0.95,0.89,0.96\}\,,\\
& \frac{\sigma_{\rm P+B}}{\sigma_{\rm P}}
\{b_1,b_2,b_{\mathcal{G}_2},P_{\rm shot}\}
=\{0.84,0.18,0.09,0.65\}\,.
\end{split}
\ee
In general, the gain here is more modest compared to what we have obtained from the real space bispectrum.
One reason for that is the correlation
between the additional FoG counterterm $c_1$
and 
other parameters. 
% This correlation may be enough to account for $\sim 5\%$
% less improvement compared to the real space case. 
For example, the degeneracy between $c_1$ and $b_2,B_{\rm shot}$
is quite significant, which explains why the confidence intervals 
for these nuisance parameters are noticeably larger than those 
of the real space bispectrum
case. 
Another reason for the relatively 
small improvement in cosmological 
parameters is that the BAO wiggles are more suppressed in 
redshift space, c.f. Eq.~\eqref{eq:IRres}, and hence there is less 
available 
distance information.

All in all, the upshot of our analysis is that for the full 
PT challenge simulation volume the data cut for the tree-level 
redshift-space
bispectrum model is $\kmax=0.08~\hMpc$, and the addition of the 
bispectrum likelihood yields $\lesssim 10\%$ improvement on
cosmological parameters, but much larger gains on bias parameters.

\begin{figure}
\centering
\includegraphics[width=0.99\textwidth]{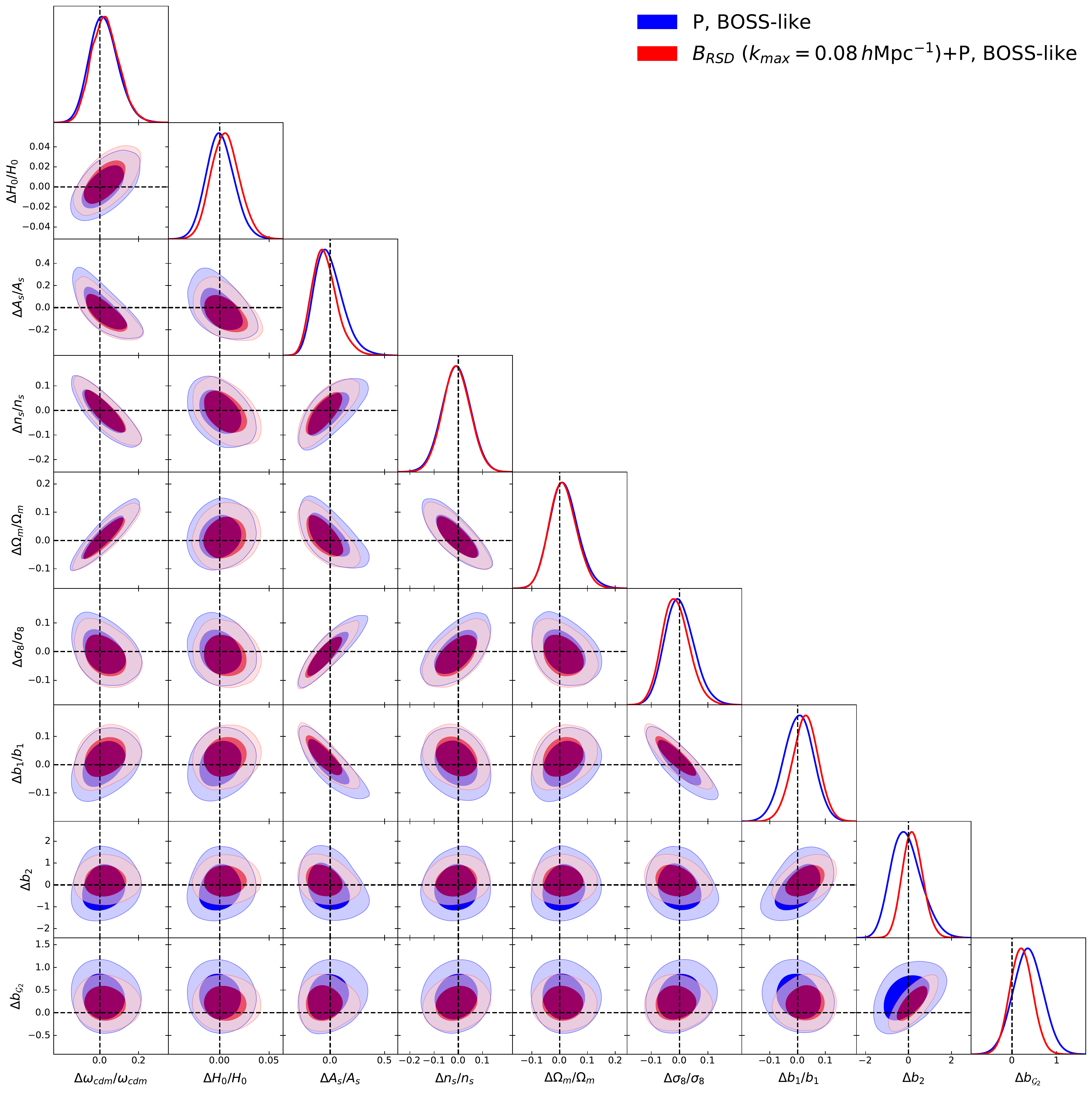}
\caption{
Same as Fig.~\ref{fig:rsd}
but with the covariance rescaled by $100$ to match 
the BOSS survey volume.
}
\label{fig:rsdBOSS}
\end{figure}

\subsection{Forecast for BOSS}
\label{sec:boss}

It is useful to re-run our analysis for the covariance that 
matches the volume of the currently available BOSS data.
In this case the covariance is larger, and hence 
we can use more aggressive data cuts
provided that the bias in cosmological parameters 
due to higher order loop corrections
is smaller than a fraction 
of the statistical error. 
In this case the power spectrum multipole analysis 
can be pushed to $k_{\rm max}=0.20~\hMpc$,
which is noticeably larger than our baseline PT challenge 
power spectrum multipole data cut 
$k_{\rm max}=0.14~\hMpc$~\cite{Chudaykin:2020ghx}.
Note that this $\kmax$ is lower than $\kmax=0.25~\hMpc$
used in Refs.~\cite{Ivanov:2019pdj,Philcox:2020vvt}
because here we include the hexadecapole moment, see Ref.~\cite{Chudaykin:2020ghx} for more detail.
Consequently, the transverse power spectrum moment $Q_0$
is taken in the range $0.2<k/(\hMpc)<0.4$~\cite{Ivanov:2021haa}.
Unfortunately, 
we cannot push 
the bispectrum analysis to 
$k_{\rm max}=0.1~\hMpc$ because the relative theory
systematic error on $\sigma_8$ there is around $3\%$. 
This is a significant 
fraction of the BOSS statistical error, $\sigma_{\sigma_8}/\sigma_8\approx 5\%$.
We have explicitly checked that the recovered value
of $\sigma_8$ is biased by $1\sigma$ of the BOSS error
when the bispectrum is taken at $k_{\rm max}=0.1~\hMpc$.
Therefore, we proceed
with the same baseline cut 
as in the PT challenge analysis of the previous 
section, $k_{\rm max}=0.08~\hMpc$.

We analyze the same PT challenge data but with 
the covariance rescaled by a factor $100$,
which is the difference between the PT challenge volume 
and the BOSS survey volume \mbox{$V_{\rm BOSS}\simeq 6$ $h^{-3}$Gpc$^3$}.
In this particular analysis, we also impose the following Gaussian prior 
on $c_1$,
\be
 c_1\sim \mathcal{N}(0,5^2)\,,
\ee
which is motivated by the EFT expectation $ c_1=\mathcal{O}(1)$.
Our results are shown in Fig.~\ref{fig:rsdBOSS} and in Table~\ref{tab:boss}.

We observe that the addition of the bispectrum 
has roughly the same impact for cosmological parameters
are the full PT challenge case, i.e.~there is 
a $\sim 10\%$ improvement on the 1d marginalized 
constraint on $A_s$ and $\sigma_8$, and barely 
any effect on other parameters. 
As far as quadratic bias parameters are concerned, 
the improvements 
for them
are less sizable. In contrast
to the full PT challenge simulation,
in the BOSS-like
power spectrum case the bias parameters are 
dominated by 
priors (given in Appendix~\ref{app:power}) and not by the data.
Hence, the power spectrum 
posteriors are narrower to begin with.
Still, 
the addition of the bispectrum data 
sharpens $b_2$ and $b_{\mathcal{G}_2}$
posteriors by a factor of two.

\begin{table}[!htb]
    % \caption{Global caption}
    \begin{minipage}{0.5\linewidth}
\begin{tabular}{|l|c|c|}
\hline
\multicolumn{2}{|c|}{\text{Power spectrum (PS), BOSS-like}}
\\ \hline
Parameter & 68\% limits \\ \hline
\hline

$\Delta H_0/H_0            $ & $0.001^{+0.013}_{-0.015}   $\\

$\Delta \omega_{cdm}/\omega_{cdm}$ & $0.021^{+0.065}_{-0.080}   $\\

$\Delta A_s/A_s            $ & $-0.01^{+0.10}_{-0.15}     $\\

$\Delta n_s/n_s            $ & $-0.009\pm 0.059           $\\

$\Delta b_1/b_1            $ & $0.004\pm 0.053            $\\

$\Delta b_2                $ & $-0.09^{+0.59}_{-0.80}     $\\

$\Delta b_{\mathcal{G}_2}  $ & $0.36\pm 0.33              $\\
{$P_{{shot} }    $} & $-0.13\pm 0.51             $\\
\hline
$\Delta \sigma_8/\sigma_8  $ & $0.000^{+0.046}_{-0.055}   $\\

$\Delta \Omega_m/\Omega_m  $ & $0.016^{+0.046}_{-0.055}   $\\
\hline 
\end{tabular}
\end{minipage}
\begin{minipage}{0.5\linewidth}
\begin{tabular}{|l|c|c|}
\hline
\multicolumn{2}{|c|}{\text{PS + bispectrum, BOSS-like}}
\\ \hline
Parameter & 68\% limits \\ \hline
\hline

$\Delta H_0/H_0            $ & $0.006^{+0.013}_{-0.015}   $\\

$\Delta \omega_{cdm}/\omega_{cdm}$ & $0.029^{+0.063}_{-0.080}   $\\

$\Delta A_s/A_s            $ & $-0.049^{+0.093}_{-0.13}   $\\

$\Delta n_s/n_s            $ & $-0.008\pm 0.057           $\\

$\Delta b_1/b_1            $ & $0.028\pm 0.047            $\\

$\Delta b_2                $ & $0.19^{+0.44}_{-0.50}      $\\

$\Delta b_{\mathcal{G}_2}  $ & $0.21\pm 0.25              $\\

$P_{{shot} }    $ & $-0.38\pm 0.40             $\\

$B_{{shot} }    $ & $1.23\pm 0.89              $\\

$c_1                       $ & $0.1\pm 4.8                $\\

\hline

$\Delta \sigma_8/\sigma_8  $ & $-0.013^{+0.044}_{-0.052}  $\\

$\Delta \Omega_m/\Omega_m  $ & $0.012^{+0.045}_{-0.052}   $\\

\hline

\end{tabular}
\end{minipage}
\caption{\label{tab:boss}
1d marginalized limits from analyses of the redshift space galaxy power spectrum (left panel)
and the joint power and bispectrum data (right panel) from the PT challenge simulation with the covariance rescaled 
to match the volume of the BOSS survey, as shown in Fig.\,\ref{fig:rsdBOSS}.
}
\end{table}

\section{Comparison with Previous Work}
\label{sec:compar}

Our analysis complements and extends other works on the 
galaxy bispectrum. Therefore, it is useful to
compare our study with the most relevant literature.

Ref.~\cite{Oddo:2019run} studied the real space halo galaxy bispectrum
from simulations with the overall volume similar to that of the 
PT challenge suite. This work used the tree-level bispectrum model
to fit the pure bispectrum data 
(in the absence of the power spectrum), and has established 
that this model works up to $\kmax=0.082~\hMpc$, in agreement 
with our baseline result $\kmax=0.08~\hMpc$. This work did not find any 
significant deviations from Poissonian sampling for the halo bispectrum. 
In contrast, we did find the sub-Poissonian shot noise
for the PT challenge galaxies. 
Importantly, this detection 
is 
driven by the power spectrum data,
which yields a $\gtrsim 2\sigma$ deviation from the Poissonian sampling 
even in the absence of any bispectrum data.
This can be compared with the bispectrum data alone (see Fig.~\ref{fig:rsdONLY}),
which is not precise enough to constrain the shot noise.  
When the two likelihood are combined, we obtain much tighter constraints on the shot noise parameters
than the bispectrum alone, which explains 
why our analysis 
is more sensitive to shot noise corrections
than that of Ref.~\cite{Oddo:2019run}.
Nevertheless, our result is not surprising, given that on general grounds 
we do expect halo stochasticity to be different from 
that of galaxies, see e.g.~\cite{Schmittfull:2020trd}.
The importance of beyond-Poissonian sampling for 
primordial non-Gaussianity constraints from the bispectrum
was also emphasized in~\cite{MoradinezhadDizgah:2020whw}.

Ref.~\cite{Eggemeier:2021cam} presented constraints on the galaxy bias 
parameters from the combination of the real space power spectrum
and bispectrum data. This work used a one-loop theoretical error model
for the bispectrum, which allowed one to push the analysis to small
scales and achieve parameter measurement precision 
similar to ours while using smaller effective volume $V_{\rm eff}=6~h^{-3}$
Gpc$^3$. An important observation is our analysis confirms 
the result of Ref.~\cite{Eggemeier:2021cam} that the quadratic
bias parameters of BOSS-like galaxies do not follow 
halo-calibrated dependencies on linear bias $b_1$. The deviations
from these dependencies that we find in our work agree very well with
those reported in Ref.~\cite{Eggemeier:2021cam}.

It is also worth comparing conclusions on the cosmological parameter 
improvements from the bispectrum in real space from Ref.~\cite{Eggemeier:2021cam}
and from our redshift space analysis.
Ref.~\cite{Eggemeier:2021cam} showed that constraints on 
$A_s$ typically improve by factors of $4-6$ in real space.
This improvement factor stays roughly the same regardless 
of whether the tree-level or the one-loop bispectrum model is used. 
In contrast to this, our analysis implies that the bispectrum monopole
sharpens the $A_s$ constraints only by $\sim 20\%$
in redshift space. This happens because the notorious degeneracy 
between the linear bias $b_1$ 
and $A_s$, which plagues real space 
analyses, is lifted in redshift space already at the level of power 
spectrum multipoles.

The bispectrum monopole 
of BOSS-like mocks and the actual 
bispectrum data from the CMASS north galactic cap (NGC) sample 
were analyzed in Ref.~\cite{DAmico:2019fhj}.
This analysis is closest to ours since it uses essentially
a similar EFT theoretical model for the power spectrum part. 
However, its bispectrum analysis 
is different from ours by a number of instances. 
First, systematic errors in the window function
treatment forced Ref.~\cite{DAmico:2019fhj} 
to discard low wavenumber modes, i.e.~use $k_{\rm min}=0.04~\hMpc$.\footnote{In principle, this issue can be avoided with the help of unwinowed estimators
implemented along the lines of Refs.~\cite{Philcox:2020vbm,Philcox:2021ukg}.}
Second, similarly to us, the authors of
Ref.~\cite{DAmico:2019fhj} used the tree-level EFT model for the bispectrum
monopole.
However, they ignored IR resummation (which is necessary 
already at the tree level~\cite{Blas:2016sfa,Baldauf:2016sjb}) 
and additional corrections
due to FoG and binning, which was partly justified by 
the smallness of the total simulation volume of that work
compared to ours.
Nevertheless, the final scale cuts
$\kmax=(0.08-0.1)~\hMpc$
of the bispectrum analysis of Ref.~\cite{DAmico:2019fhj}
are consistent with our choice $\kmax=0.08~\hMpc$.
Finally, Ref.~\cite{DAmico:2019fhj} found 
that the bispectrum data from one BOSS data chunk (CMASS NGC) 
sharpens the constrain on
$A_s$ by $\lesssim 20\%$ and leaves intact other cosmological 
parameters. We have found a quantitatively similar behavior in our analysis, 
see Fig.~\ref{fig:rsdBOSS}. It will be interesting to see 
how much the constraints improve in the 
analysis of the actual BOSS data 
with our likelihood. We leave this for future work~\cite{future}.

Finally, it is worth comparing our results with those 
from the MCMC forecast for the Euclid-like 
survey from Ref.~\cite{Chudaykin:2019ock}.
This work used a very similar methodology 
and found that the addition of the tree-level 
bispectrum monopole likelihood leads to $\sim (10-50)\%$
on all relevant cosmological parameters of the
 $\Lambda$CDM model with massive neutrinos.
Our analysis is different from Ref.~\cite{Chudaykin:2019ock} in 
several aspects. First, unlike Ref.~\cite{Chudaykin:2019ock}, 
our baseline power spectrum 
likelihood contains the real space power spectrum $Q_0$~\cite{Ivanov:2021haa}.
Moreover, our likelihood here
includes physical priors on nuisance parameters, whereas 
Ref.~\cite{Chudaykin:2019ock} did not assume any priors on
them. These two factors may 
diminish relative information content of the 
bispectrum in our work. 
Second, we analyze only one redshift bin here, whereas 
Ref.~\cite{Chudaykin:2019ock} considers a more realistic 
data sample spread across 8 different bins. 
Clearly, this latter case contains more distance information
that can be extracted 
through the AP effect.
Third, we impose the BBN prior on $\Omega_b$ here, while
Ref.~\cite{Chudaykin:2019ock} fits this parameter directly from the 
large-scale structure data.
Despite these significant differences, 
one observes a qualitative agreement between our results: 
in both cases the tree-level 
bispectrum monopole improves cosmological 
parameter constrains by tens of percent.

\section{Conclusions}
\label{sec:concl}

In this work we have studied the
cosmological information present in the redshift-space bispectrum
monopole of PT challenge simulation galaxies. 
We analyze the joint power spectrum and bispectrum 
likelihood using the one-loop EFT model for the power 
spectrum and the tree-level model for the bispectrum.
This is a fully consistent approach as for both 
statistics we use the perturbative density field expanded to 
third order in the linear solution. 
Our bispectrum theoretical templates include, 
for the first time, all the effects that are needed to 
describe the data at this order:
tree-level 
IR resummation, corrections due to
discreteness, FoG, and the 
AP
effect.
Our main results are 
\begin{itemize}
	\item The tree-level bispectrum model is valid 
	up to $\kmax=0.08~\hMpc$ for a BOSS-like luminous red galaxy sample.
	\item The addition of the tree-level bispectrum 
	likelihood to the power spectrum one
	leads to moderate improvements of constraints 
	on cosmological parameters by $\lesssim 10\%$.
	\item The improvement on bias parameters is very significant. 
	The errorbars on the quadratic local in density bias  
	$b_2$ and the tidal bias $b_{\mathcal{G}_2}$
	shrink by more than a factor of 10
	after adding the bispectrum data.
	\item 
	 We have found that
	the quadratic galaxy bias parameters 
	are quite different from
	biases of host dark matter halos.
	This confirms the trend seen in the literature~\cite{Eggemeier:2021cam,Barreira:2021ukk}. 
\end{itemize}

On the technical side, we have proposed a new 
efficient approach to account for binning effects
by a combination of the integral approximation 
and discreteness weights,
and also studied in detail 
the dependence of our results  
on bispectrum covariance matrix choices.
% We have created likelihoods and infrastructure 
% for this analysis.

There are several ways to extend our analysis.
First, it would be important to 
upgrade our theory model with the redshift-space 
one-loop bispectrum calculations. 
In particular, we have found 
that at $\kmax>0.08~\hMpc$ the data 
shows evidence for FoG,
which is a loop effect in the EFT nomenclature.
Given that the one-loop calculation significantly 
extends the regime of validity of the EFT 
in the power spectrum case, one may expect that 
a similar improvement can take place for the 
bispectrum. It is important to notice that for 
consistency one needs to compute the power spectrum
at two loop order when considering the one-loop
bispectrum.

Moreover, it is also interesting to consider higher angular
moments of the redshift-space bispectrum. 
Various forecasts suggest that these moments
may
contain significant cosmological information, see 
e.g.~\cite{Yankelevich:2018uaz}. We plan 
to verify these results 
in an actual analysis of simulated or real data.
Importantly, higher order
bispectrum multipoles are sensitive to 
FoG, and hence one-loop 
corrections are desirable 
for their systematic study.
This issue can be mitigated with an analog of the 
transverse moment $Q_0$
for the bispectrum. 
We plan to study this statistics in future. 

Another natural step is the analysis of the actual bispectum data from the BOSS survey~\cite{future}. Our work suggests that the inclusion of the bispectrum may improve constraints on the mass fluctuation
amplitude by $\sim 10\%$.
This improvement is not very dramatic, but it should be 
pointed out that so far we have considered only 
the minimal $\Lambda$CDM model.
The information content of the redshift space bispectrum 
can be richer in nonminimal cosmological models, which may have some
implications for certain tensions, e.g.~the so-called $\sigma_8$ tension~\cite{DiValentino:2020vvd}. 

Finally, it would be interesting to 
repeat our analysis for the emission line galaxies,
which will be the main targets of future surveys 
like DESI~\cite{Aghamousa:2016zmz} and Euclid~\cite{Laureijs:2011gra,Amendola:2016saw}. 
Emission line galaxies are less biased than the red luminous galaxies whose mocks we studied in this paper.
Moreover, recent measurements suggest that 
they are less affected by FoG~\cite{deMattia:2020fkb,Ivanov:2021zmi}, which implies 
that the EFT model may perform better for this sample.

\paragraph{Note added.} While this paper was being 
prepared, a new work \cite{Oddo:2021iwq}
studying the information content 
of the real space bispectrum 
appeared. When results overlap, they are in agreement.

\paragraph{Acknowledgments.} We thank Giovanni Cabass, Roman Scoccimarro 
and Jay Wadekar for fruitful discussions. 
The work of 
MMI has been supported by NASA through the NASA Hubble Fellowship grant \#HST-HF2-51483.001-A awarded by the Space Telescope Science Institute, which is operated by the Association of Universities for Research in Astronomy, Incorporated, under NASA contract NAS5-26555.
OHEP thanks the Simons Foundation for support.
This work was supported in part by MEXT/JSPS KAKENHI Grant Number JP19H00677, JP20H05861 and JP21H01081.
We also acknowledge financial support from Japan Science and Technology Agency (JST) AIP Acceleration Research Grant Number JP20317829.
The simulation data analysis was performed partly on Cray XC50 at Center for Computational Astrophysics, National Astronomical Observatory of Japan.

\appendix 

\section{Tests of Binning}
\label{app:bins}

In order to account for binning effects, one should ideally evaluate the full sum over all possible triangle 
configurations inside the bin.
However, this evaluation is computationally very 
expensive. In the main text we have used the integral approximation
along with discreteness weights that correct for the inaccuracy 
of this approximation for the folded triangles.
In this appendix we present an alternative to this scheme,
which works well for the real space bispectrum. 

The main goal of our binning scheme 
is to generate many ``fundamental'' triangle configurations
based on the true $k_f$ and then sum them into appropriate bins.
It is computationally expensive 
to generate all the fundamental triangles
on the actual 3d Fourier grid. Moreover, it is also 
not efficient, because the fundamental grid contains
a large number of identical fundamental triangles.
We can avoid that by organizing a sum 
over \textit{unique} fundamental triangle configurations
with $q_1\geq q_2\geq q_3$, where $q_i$ are wavevector moduli
of fundamental triangles.
In this case we need to sample the bispectra over 
a relatively small grid of wavenumbers.
We will call this method ``approximate 1d binning'' in what follows.
It is based, essentially, on taking the 
integral expression Eq.~\eqref{eq:Bint2}
and approximating it with a sum over 
appropriate
discrete configurations of wavevector moduli.

Let us start with the integral approximation Eq.~\eqref{eq:Bint2},
obtained after eliminating most of angular variables
by means of the Dirac delta-function.
In real space the integrals over $\mu$ and $\phi$
drop out of this expression because the bispectrum 
does not depend on angles.
Now we can write down the following discrete approximation
to the final integral,
\be
\label{eq:Approx1p5}
\begin{split}
\hat{B}^{\rm int}_0\simeq 
\frac{V\sum_{123} 
q_1q_2q_3 B(q_1,q_2,q_3)}{V\sum_{123} 
q_1q_2q_3~1}
\,,
\end{split}
\ee
where the sum $\sum_{123}$ is taken
over all configurations of momentum moduli $q_i$ that fall in the bin.
This sum contains many indistinguishable modes.
Now we replace this sum with a discrete sum over 
independent triangle configurations only,
\be
\label{eq:Approx2}
\begin{split}
% & \prod_{i=1}^3\sum_{\q_i\in k_i}\delta_K(\q_{123})B(q_1,q_2,q_3)\\
\hat{B}^{\rm int}_0\simeq 
\frac{V\sum_T
q_1q_2q_3 B(q_1,q_2,q_3)}{V\sum_T
q_1q_2q_3~1}
% \Bigg|_{q_3\leq k_2+\Delta k/2}
% _{q_3\geq k_2-\Delta k/2}
\,,
\end{split}
\ee
where the sum $\sum_T$ runs over all unique triangles
that fall in the bin $(k_1,k_2,k_3)$ and that
respect
the $k_f$ spacing,
\be
\label{eq:sumT}
\sum_T\equiv 
\sum_{q_1=\text{max}(k_f,k_{1}-\Delta k/2)}^{k_{1}+\Delta k/2}
\sum_{q_2=\text{max}(k_f,k_2-\Delta k/2)}^{\text{min}(k_2+\Delta k/2,q_1)}
% \Bigg|_{q_2\leq k_2+\Delta k/2}
\sum_{q_3=\text{max}(k_{f},k_3-\Delta k/2,q_1-q_2)}^{\text{min}(k_3+\Delta k/2,q_2)} \,.
\ee
To compute the sum in Eq.~\eqref{eq:Approx2}
in practice, we generate a grid of 
tuples ($q_1,q_2,q_3$) with spacing $\Delta q=k_f$
and select only those that satisfy the constraints 
of Eq.~\eqref{eq:sumT} for each bin ($k_1,k_2,k_3$).
Notice that we have used the isotropy of the bispectrum 
in our derivation, which is certainty not true
in redshift space. We apply the approximate 1d binning
scheme in real space only.

\begin{figure}
\centering
\includegraphics[width=0.99\textwidth]{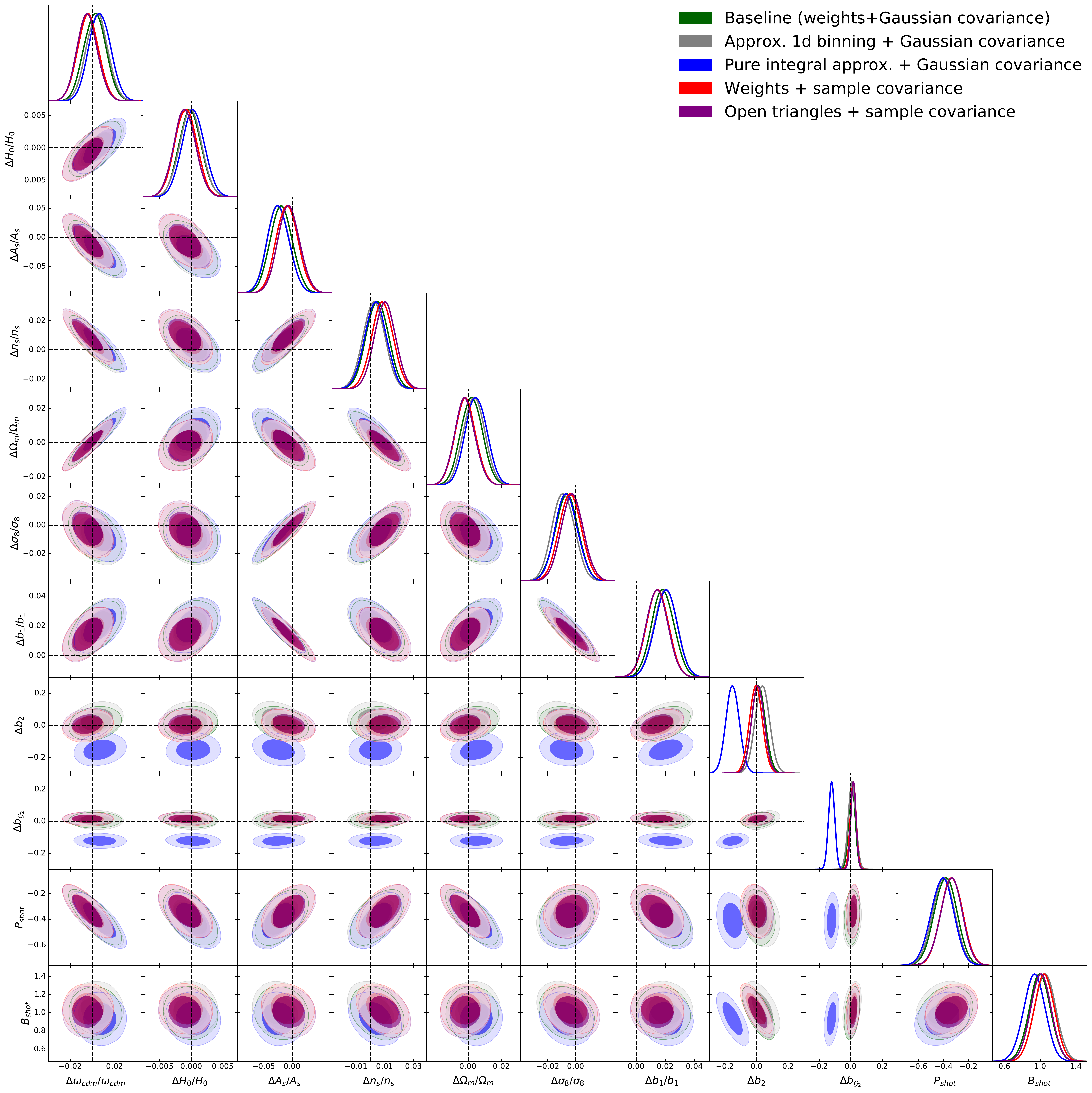}
\caption{Triangle plots and 1d marginalized posteriors of cosmological 
and nuisance parameters from the following analyses that differ
by the real space bispectrum likelihood treatments: 
baseline (Gaussian covariance + discretness weights),
approximate 1d binning,
integral approximation for binning 
without additional binning corrections (+ Gaussian covariance for the last 
two cases); likelihood based on the bispectrum sample covariance 
(discreteness weights), and the likelihood that includes 
the extra open triangles (+ sample covariance).
}
\label{fig:covratlike}
\end{figure}

Eq.~\eqref{eq:Approx2} is not exact
because it was derived from the integral expression~\eqref{eq:Bint2},
which is approximate on its own.
But we can still use it 
as an alternative prescription for the binning effects
that will allow us to assess the systematic error 
of our baseline discreteness weight method.
The two methods can be compared in Fig.~\ref{fig:covratlike}, 
``Baseline'' vs. ``Approx 1d binning.''
We can clearly see that they yield almost identical results.
This validates our discreteness weight approach adopted in the main 
analysis. In this plot, we also show results from
the ``Pure integral approximation'' obtained from
the bispectrum model with
the integral approximation but
without discreteness weights or any additional corrections. 
This prescription
leads to significant biases in the recovery of
quadratic bias parameter $b_2$ and $b_{\mathcal{G}_2}$.

\section{Impact of Open Triangles}
\label{app:coll}
In principle, we could also include in our analysis the open triangles, i.e.~the triangle bins that do not satisfy 
$|k_3-k_2|<k_1<k_3+k_2$ at their centers.
We refrained from doing so because of the reasons listed in the main text.
In this Appendix we explicitly check that neglecting these triangles 
does not lead to any appreciable loss of information.
We include these triangles in the analysis by adopting 
the approximate 1d binning scheme described above.
We have found that the Gaussian covariance approximation is 
very inaccurate for them, and therefore use a diagonal sample covariance
matrix in our likelihood. 
The sample covariance 
matrix approximation for ``usual'' closed triangle configurations
is validated in the next section, showing that it 
leads to essentially the same results as our baseline Gaussian covariances.

The results of our analysis of the bispectrum likelihood including 
open triangles are shown in Fig.~\ref{fig:covratlike},
which should be compared with the case ``Weights + sample covariance.''
 We see that the posterior
distribution in this case is almost identical to that of the usual sample covariance
analysis without open triangles, which implies that they can safely neglected for the purposes
of this paper.

\section{Covariance Matrix Tests}
\label{app:covar}

\begin{figure}
\centering
\includegraphics[width=0.99\textwidth]{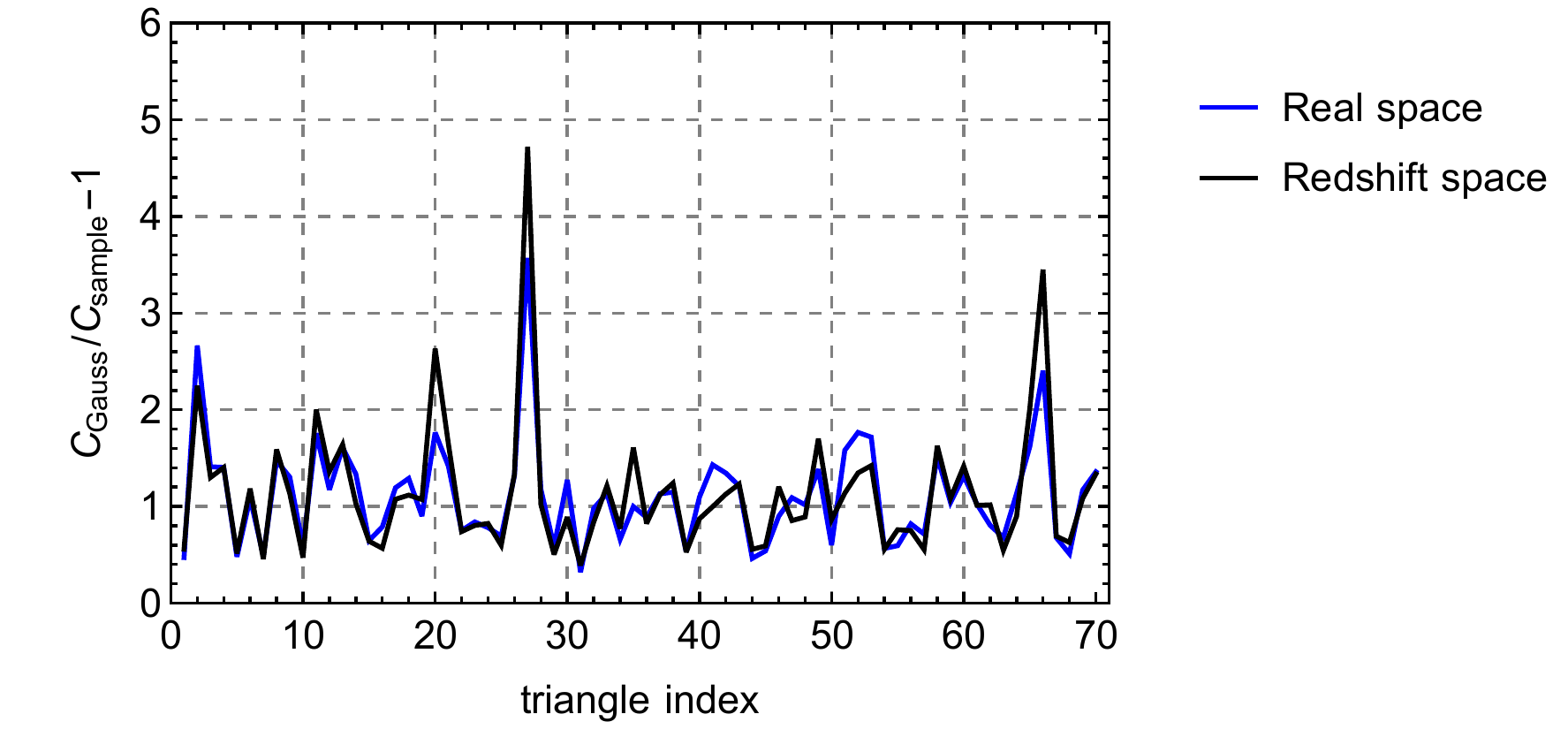}
\caption{Ratio of diagonal elements of the bispectrum covariance matrix
computed in the Gaussian approximation and the sample covariance
extracted from 10 PT challenge simulation boxes.
}
\label{fig:covrat}
\end{figure}

To test our baseline Gaussian covariance model, in this section
we run our analysis with bispectrum likelihood based on sample covariance 
matrix estimators. The PT challenge suite consists of 10 boxes only, which means 
that the relative error on elements of the sample covariance in this case is 
around ${10}^{-1/2}\sim 30\%$. Since the sample covariance is not invertible 
for our baseline bispectrum data with 70 triangle bins, 
we will use only its diagonal part. 
This should still be a good approximation 
on large scales where the bispectrum
covariance is dominated by the Gaussian diagonal 
contribution.
The elements of our sample covariance normalized to the 
predictions of the Gaussian approximation are shown in Fig.~\ref{fig:covrat}.
We see that the ratio is scattered around unity with most of the points dispersed 
within $\sim 50\%$ in accordance with the expected variance. However, we also 
observed several notable outliers. Nevertheless, the posterior distribution 
from the likelihood based on the sample covariance
is almost identical to that of the baseline analysis, 
see Fig.~\ref{fig:covratlike} for the real space case 
and Fig.~\ref{fig:covratlike2} for the redshift space case. We see that the main effect of the sample 
covariance is to shift the posterior distributions, but these shifts 
do not exceed $1\sigma$, which is an 
expected effect of the sampling noise in the covariance~\cite{Philcox:2020zyp,Wadekar:2020hax}.

\subsection{Theoretical error and cross-covariance}

\begin{figure}
\centering
\includegraphics[width=0.99\textwidth]{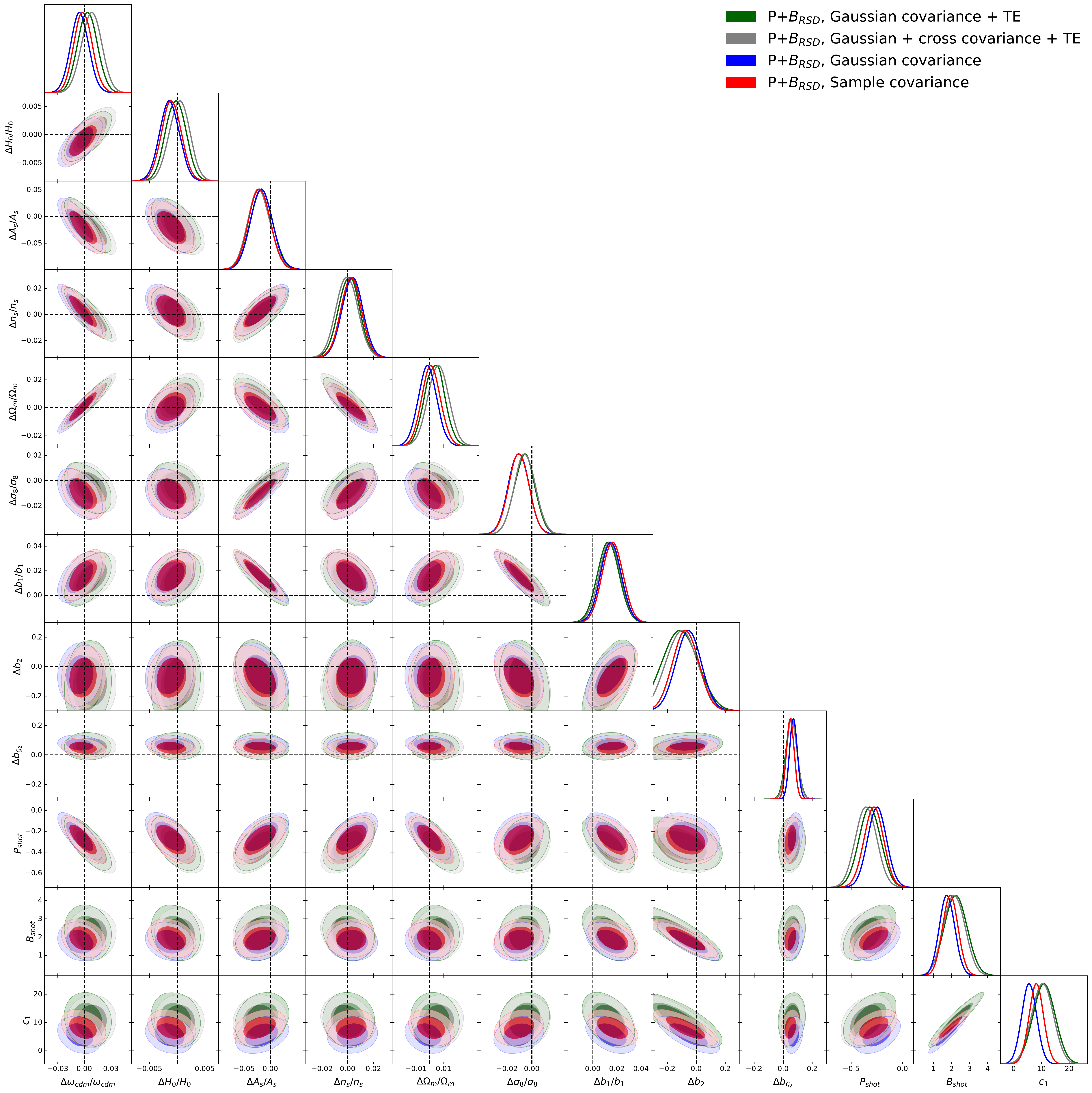}
\caption{Triangle plots and 1d marginalized posteriors of cosmological 
and nuisance parameters from the joint redshift-space power spectrum
and bispectrum  
likelihoods built with different covariance matrices:
the Gaussian covariance,
the Gaussian covariance including the theoretical error (TE),
the Gaussian covariance including the theoretical error and the cross-covariance 
between the power spectrum and bispectrum,
and the bispectrum sample covariance.
In all cases we have used the discreteness weights and $\kmax=0.08~\hMpc$
for the bispectrum likelihood.
}
\label{fig:covratlike2}
\end{figure}

\begin{table}[!htb]
    % \caption{Global caption}
    \begin{minipage}{0.5\linewidth}
\begin{tabular}{|l|c|c|}
\hline
\multicolumn{2}{|c|}{\text{TE covariance}}
\\ \hline
Parameter & 68\% limits \\ \hline
\hline

$\Delta H_0/H_0            $ & $-0.0002\pm 0.0018         $\\

$\Delta \omega_{cdm}/\omega_{cdm}$ & $0.004\pm 0.011            $\\

$\Delta A_s/A_s            $ & $-0.017\pm 0.021           $\\

$\Delta n_s/n_s            $ & $0.0009\pm 0.0082          $\\

$\Delta b_1/b_1            $ & $0.0128\pm 0.0089          $\\

$\Delta b_2                $ & $-0.11\pm 0.12             $\\

$\Delta b_{\mathcal{G}_2}  $ & $0.058\pm 0.038            $\\

{$P_{{shot} }    $} & $-0.32\pm 0.11             $\\

{$B_{{shot} }    $} & $2.36\pm 0.64              $\\

$c_1                       $ & $11.0\pm 4.0               $\\

\hline 
$\Delta \sigma_8/\sigma_8  $ & $-0.0059\pm 0.0083         $\\

$\Delta \Omega_m/\Omega_m  $ & $0.0035\pm 0.0072          $\\

\hline 
\end{tabular}
\end{minipage}
\begin{minipage}{0.5\linewidth}
\begin{tabular}{|l|c|c|}
\hline
\multicolumn{2}{|c|}{\text{TE + cross covariance}}
\\ \hline
Parameter & 68\% limits \\ \hline
\hline

$\Delta H_0/H_0            $ & $0.0004\pm 0.0018          $\\

$\Delta \omega_{cdm}/\omega_{cdm}$ & $0.008\pm 0.011            $\\

$\Delta A_s/A_s            $ & $-0.021\pm 0.021           $\\

$\Delta n_s/n_s            $ & $-0.0006\pm 0.0084         $\\

$\Delta b_1/b_1            $ & $0.0140\pm 0.0088          $\\

$\Delta b_2                $ & $-0.10\pm 0.11             $\\

$\Delta b_{\mathcal{G}_2}  $ & $0.059\pm 0.037            $\\

{$P_{{shot} }    $} & $-0.35\pm 0.10             $\\

{$B_{{shot} }    $} & $2.15\pm 0.59              $\\

$c_1                       $ & $10.4\pm 3.8               $\\

\hline 

$\Delta \sigma_8/\sigma_8  $ & $-0.0056\pm 0.0084         $\\

$\Delta \Omega_m/\Omega_m  $ & $0.0061\pm 0.0075          $\\

\hline 
\end{tabular}
\end{minipage}
\caption{\label{tab:cross}
1d marginalized limits from analyses 
of the redshift space bispectrum monopole data at $\kmax=0.08~\hMpc$
with two additional ingredients: the one-loop theoretical error (TE) bispectrum 
covariance
(left table) and the TE bispectrum 
covariance plus the cross-covariance between the power spectrum multipoles
and the bispectrum (right table).
}
\end{table}

We additionally check the stability of our results w.r.t.~the inclusion
of the theoretical error covariance to the bispectrum
and the cross-covariance between the power spectrum multipoles 
and the bispectrum monopole.

The theoretical error covariance accounts for the imperfectness 
of the particular theoretical model that is used to fit the data. 
In the EFT approach theoretical calculations are done up to a fixed 
order on scales where higher order corrections are estimated to be 
negligible. A more systematic way to account for these corrections is 
to marginalize over their approximate shape 
dictated by the EFT power counting~\cite{Baldauf:2016sjb,Chudaykin:2020hbf}.
This marginalization leads to a simple 
change of the covariance matrix by an additive correlated contribution. 
We incorporate the theoretical error covariance for the bispectrum 
following Ref.~\cite{Baldauf:2016sjb}.
We use the following one-loop bispectrum theoretical error kernel 
\be
\label{eq:EB}
E_B(k_1,k_2,k_3)=B_{\rm tree}(k_1,k_2,k_3,z)
D^2_+(z)\left(\frac{k_1+k_2+k_3}{3\times 0.23~\hMpc} \right)^{3.3}\,,
\ee
whose amplitude is reduced by a factor of $3$ compared to 
Ref.~\cite{Baldauf:2016sjb}. We do so because the original envelope 
of Ref.~\cite{Baldauf:2016sjb} was calibrated to one-loop calculations 
at $k\sim 0.2~\hMpc$ which is larger than our baseline cut $\kmax=0.08~\hMpc$.
We have checked that on these scales the original theory error kernel of 
Ref.~\cite{Baldauf:2016sjb} overestimates the actual size of one-loop 
matter bispectrum corrections, and therefore have accounted for it by 
multiplying this kernel by a factor $1/3$. Using Eq.~\eqref{eq:EB},
the theoretical error covariance can be written as 
\be
C^{B~(\rm TE)}_{TT'}= E_B(k_1,k_2,k_3)E_B(k'_1,k'_2,k'_3)\prod_{i=1}^3\e^{-\frac{(k_i-k'_i)}{2\delta k^2}}\,,
\ee
where the coherence scale $\delta k=0.1~\hMpc$ following Refs.~\cite{Chudaykin:2019ock,Chudaykin:2020hbf}. The full covariance is given by 
\be
 C^{B}_{TT'}=C^{B~(\rm Gauss)}_{TT'} + C^{B~(\rm TE)}_{TT'}\,.
\ee
The result of our analysis of the bispectrum likelihood with the theoretical 
error covariance are presented in Fig.~\ref{fig:covratlike2}
and in Table~\ref{tab:cross}.
We see the inclusion of the theoretical error covariance
leads to a moderate inflation of errorbars and insignificant 
shifts of some posteriors. 

Finally, we include the cross-covariance between the power spectrum multipoles 
and the bispectrum monopole 
in our likelihood. 
We compute this cross-covariance in the tree-level approximation
along the lines of Ref.~\cite{Sefusatti:2006pa}, see Appendix~\ref{app:acovar}
for more detail. The results of this analysis are displayed in 
the same
Fig.~\ref{fig:covratlike2} and Table~\ref{tab:cross}.
The impact of the cross-covariance is quite marginal -- the posteriors
are virtually identical to those of the previous analysis which treated 
the bispectrum and the power spectrum uncorrelated. This is consistent with 
common expectations that the cross-covariance is negligible on large scales~\cite{Baldauf:2016sjb,Chudaykin:2019ock}.

All in all, the analyses that we have carried out 
suggest that our baseline results are stable 
w.r.t.~the choice of covariance matrices.

\section{Baseline power spectrum likelihood}
\label{app:power}

Our baseline power spectrum likelihood consists of two pieces: 

\paragraph{Redshift space multipoles}$\ell=0,2,4$
with $\kmax=0.14~\hMpc$. We build the likelihood using 
the Gaussian approximation for the covariance matrix of these multipole
moments.
In the previous work~\cite{Nishimichi:2020tvu}
we have checked that the one-loop EFT model provides 
an accurate and unbiased fit to the data in this range.

\paragraph{Transverse moment (real space) power spectrum} in the range $0.14~\hMpc<k<~0.4~\hMpc$. We use the Gaussian
covariance for real space part of the power spectrum likelihood.

Our power spectrum likelihood depends on
the following nuisance parameters~
\be
\{b_1,b_2,b_{\mathcal{G}_2},b_{\Gamma_3},c_0,c_2,b_4,a_0,a_2,P_{\rm shot}\}\,, 
\ee
for which we assume following physically-motivated priors~\cite{Chudaykin:2020aoj,Chudaykin:2020hbf,Chudaykin:2020ghx,Ivanov:2021haa}:
\be
\begin{split}
& b_1\in (1,4)\,,\quad b_2\sim \mathcal{N}(0,1^2)\,,\quad b_{\mathcal{G}_2}\sim \mathcal{N}(0,1^2)\,,\quad b_{\Gamma_3}\sim \mathcal{N}\left(\frac{23}{42} (b_1^{\rm fid}-1),1^2\right)\,,\\
& \frac{c_0}{(h^{-1}\text{Mpc})^2} \sim \mathcal{N}(4,10^2)\,,\quad \frac{c_0}{(h^{-1}\text{Mpc})^2} \sim \mathcal{N}(20,20^2)
\,,\quad \frac{c_4}{(h^{-1}\text{Mpc})^2} \sim \mathcal{N}(-10,20^2)
\,,\\
& \frac{b_4}{(h^{-1}\text{Mpc})^4} \sim \mathcal{N}(500,500^2)\,,\quad 
a_0  \sim \mathcal{N}(0,1^2)\,,\quad a_2  \sim \mathcal{N}(0,1^2)\,,\quad P_{\rm shot} \sim \mathcal{N}(0,1^2)\,.
\end{split} 
\ee
Note that we use the following parametrization for the stochastic part of the 
redshift-space power spectrum,
\be
P_{\rm stoch}(k,\mu)=\frac{1+P_{\rm shot}}{\bar n} + (a_0+a_2\mu^2)
\left(\frac{k}{0.45~\hMpc} \right)^2 \frac{1}{\bar n}\,.
\ee

\section{Real space power spectrum + bispectrum 
analysis}
\label{app:real}

In this appendix 
we study the information content 
of clustering statistics purely in real space.
We analyze the real space 
galaxy power spectrum of PT challenge simulation
at $\kmax=0.2~\hMpc$
and the real space bispectrum at $\kmax=0.08~\hMpc$.
The real space power spectrum case is very different 
from the redshift space one. In the absence 
of RSD the degeneracy between the linear 
galaxy bias $b_1$
and clustering amplitude $\sigma_8$
is largely unbroken. 
Moreover, the real space case 
does not capture the distance information,
which should result in larger errorbars on $H_0$.

\begin{figure}
\centering
\includegraphics[width=0.99\textwidth]{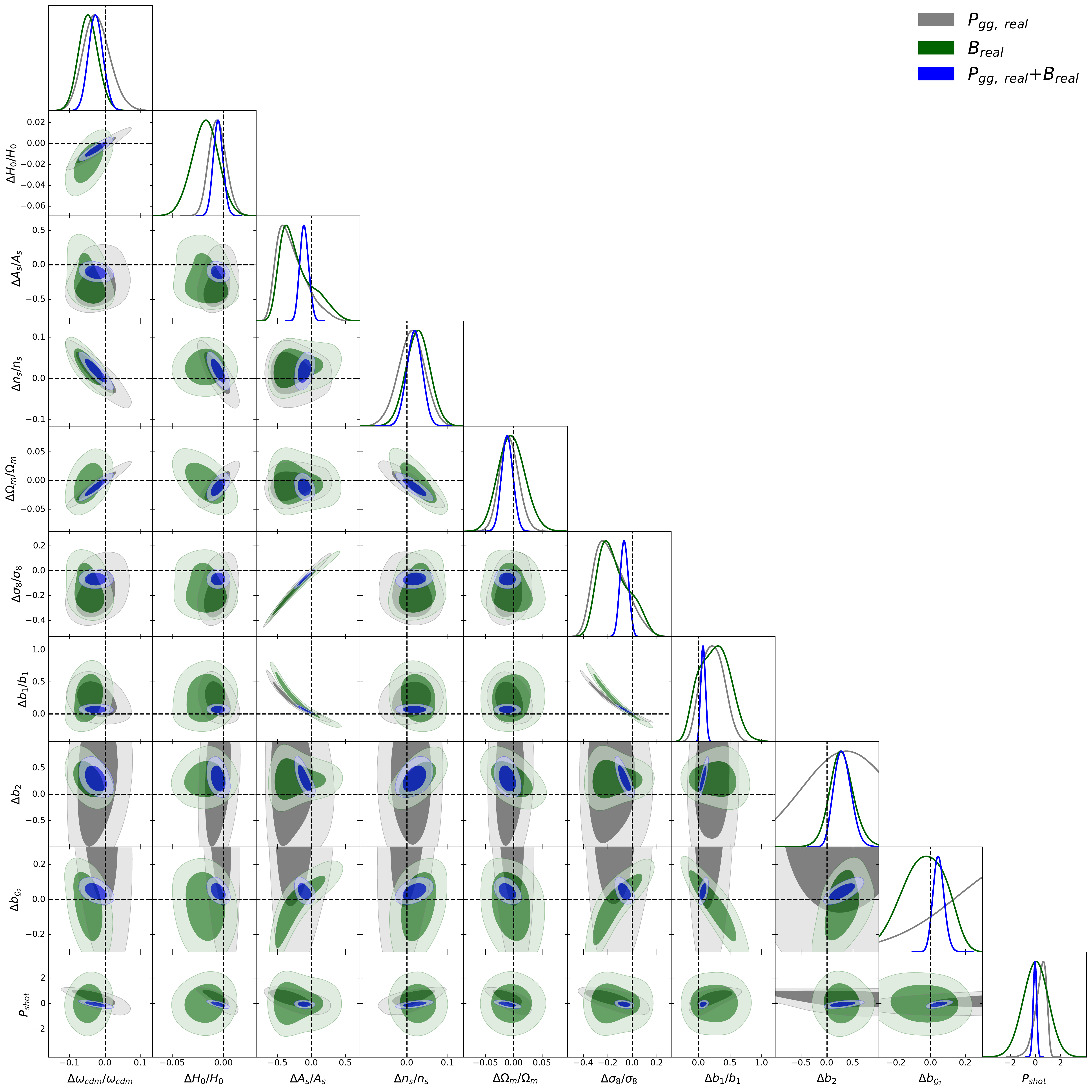}
\caption{Triangle plots and 1d marginalized posteriors of cosmological 
and nuisance parameters from the real space 
power spectra and bispectra data of the PT challenge
simulation. 
For compactness, only 
linear and quadratic bias parameters are
shown.
}
\label{fig:realall}
\end{figure}

Our results are shown in Fig.\ref{fig:realall} and in table~\ref{tab:realps}.
The real space power spectrum data $P_{\text{gg, real}}$
is much less constraining than the dataset $[P_0,P_2,P_4,Q_0]$
that we are using in our baseline redshift space 
power spectrum analysis. 
In particular, the constraints on $\omega_{\rm cdm},n_s$
and $H_0$ few times weaker, the limit on $\sigma_8$ is weaker 
by an order of magnitude.
We can also see
that the cosmological parameters' posteriors from the 
bispectrum alone are comparable 
to the power spectrum ones. 
When we combine the two statistics 
the improvement is quite significant,
e.g.~the limit on $\sigma_8$
improves by a factor of four, 
the limit on $H_0$ by $30\%$.

\begin{table}[!htb]
    % \caption{Global caption}
    \begin{minipage}{0.5\linewidth}
\begin{tabular}{|l|c|c|}
\hline
\multicolumn{2}{|c|}{$P_{\text{gg, real}}$,~\text{real space}}
\\ \hline
Parameter & 68\% limits \\ \hline
\hline
$\Delta H_0/H_0            $ & $-0.0058^{+0.0076}_{-0.0089}$\\

$\Delta \omega_{cdm}/\omega_{cdm}$ & $-0.023^{+0.033}_{-0.040}  $\\

$\Delta A_s/A_s            $ & $-0.29^{+0.13}_{-0.26}     $\\

$\Delta n_s/n_s            $ & $0.011\pm 0.033            $\\

$\Delta b_1/b_1            $ & $0.22^{+0.17}_{-0.19}      $\\

$\Delta b_2                $ & $0.43\pm 0.91              $\\

$\Delta b_{\mathcal{G}_2}  $ & $0.71^{+0.43}_{-0.55}      $\\

$P_{{shot} }    $ & $0.38^{+0.57}_{-0.29}      $\\

\hline

$\Delta \sigma_8/\sigma_8  $ & $-0.177^{+0.097}_{-0.16}   $\\

$\Delta \Omega_m/\Omega_m  $ & $-0.008^{+0.015}_{-0.018}  $\\

\hline 
\end{tabular}
\end{minipage}
\begin{minipage}{0.5\linewidth}
\begin{tabular}{|l|c|c|}
\hline
\multicolumn{2}{|c|}{$B_{\rm real}$,~\text{real space}}
\\ \hline
Parameter & 68\% limits \\ \hline
\hline
$\Delta H_0/H_0            $ & $-0.018\pm 0.013           $\\

$\Delta \omega_{cdm}/\omega_{cdm}$ & $-0.047\pm 0.028           $\\

$\Delta A_s/A_s            $ & $-0.22^{+0.13}_{-0.31}     $\\

$\Delta n_s/n_s            $ & $0.027\pm 0.029            $\\

$\Delta b_1/b_1            $ & $0.25\pm 0.24              $\\

$\Delta b_2                $ & $0.29^{+0.21}_{-0.24}      $\\

$\Delta b_{\mathcal{G}_2}  $ & $-0.04^{+0.15}_{-0.11}     $\\

{$P_{{shot} }    $} & $0.0\pm 1.0                $\\

{$B_{{shot} }    $} & $0.87\pm 0.32              $\\

\hline 

$\Delta \sigma_8/\sigma_8  $ & $-0.148^{+0.091}_{-0.16}   $\\

$\Delta \Omega_m/\Omega_m  $ & $-0.004^{+0.022}_{-0.025}  $\\

\hline 
\end{tabular}
\end{minipage}
\begin{center}
\begin{minipage}{0.5\linewidth}
\begin{tabular}{|l|c|c|}
\hline
\multicolumn{2}{|c|}{$P_{\text{gg, real}}+B_{\rm real}$,~\text{real space}}
\\ \hline
Parameter & 68\% limits \\ \hline
\hline

$\Delta H_0/H_0            $ & $-0.0054\pm 0.0047         $\\

$\Delta \omega_{cdm}/\omega_{cdm}$ & $-0.026\pm 0.020           $\\

$\Delta A_s/A_s            $ & $-0.107^{+0.060}_{-0.068}  $\\

$\Delta n_s/n_s            $ & $0.018\pm 0.019            $\\

$\Delta b_1/b_1            $ & $0.071\pm 0.036            $\\

$\Delta b_2                $ & $0.30^{+0.15}_{-0.18}      $\\

$\Delta b_{\mathcal{G}_2}  $ & $0.046^{+0.029}_{-0.034}   $\\

{ $P_{{shot} }    $} & $-0.05\pm 0.14             $\\

{ $B_{{shot} }    $} & $0.96\pm 0.18              $\\

\hline 

$\Delta \sigma_8/\sigma_8  $ & $-0.066\pm 0.033           $\\

$\Delta \Omega_m/\Omega_m  $ & $-0.011\pm 0.010           $\\

\hline 
\end{tabular}
\end{minipage}
\end{center}
\caption{\label{tab:realps}
1d marginalized limits from analyses 
of the real space power
spectrum at  $\kmax=0.20~\hMpc$
and the real space
bispectrum at $\kmax=0.08~\hMpc$.
For compactness, only 
linear and quadratic bias parameters are
shown.
}
\end{table}

\section{Power spectrum and bispectrum
covariances in perturbation theory}
\label{app:acovar}

In this Appendix we calculate tree-level covariance matrices for the power spectrum
and bispectrum. 
Let us start with the real-space estimators for the density power spectrum and bispectrum in the narrow bin approximation $\Delta k\ll k$~\cite{Scoccimarro:2015bla}
\be
\label{eq:realest}
\begin{split}
& \hat P(k_i) = \int_{q\in k_i~\text{shell}} \frac{d^3q}{(2\pi)^3{\cal N}_k }~\delta(-\q) \delta(\q)\,,\quad {\cal N}_k = 4\pi k^2 \Delta k \frac{V}{(2\pi)^3} \,. \\
& \hat B(k_1,k_2,k_3) = \prod_{i=1}^3\int_{k_i} \frac{d^3q_i}{(2\pi)^3}~\frac{(2\pi)^3\delta^{(3)}(\q_{123})}{N_T^{123}}
\delta(\q_1)\delta(\q_2)\delta(\q_3)\,,\quad N^{123}_T=8\pi^2 k_1k_2k_3\Delta k^3 \frac{V^2}{(2\pi)^6}
\,,
\end{split}
\ee 
% where we have assumed a finite bin approximation .
Using the formulas from~\cite{Mehrem:1990eg},
\be
\begin{split}
& \int r^2dr~j_0(k_1 r) j_0(k_2 r) j_0(k_3 r) =  \frac{\pi}{4}\,,\\
& \int_{k_1}\int_{k_2}\int_{k_3} [dq]^3~(2\pi)^3\delta^{(3)}(\q_{123})
=k_1k_2k_3\Delta k^3 \frac{(4\pi)^4}{(2\pi)^9}\frac{\pi}{4}\,,\\
& \delta^{(3)}_D(\k_1+\k_2+\k_3)=\frac{1}{k_1k_2k_3}\delta^{(1)}_D \l\cos(\k_1,\k_2) -\frac{k_3^2-k_1^2-k_1^2}{2k_1k_2} \r \delta^{(2)}_D \l \hat{\k}_3 - (\hat{\k}_1+\hat{\k}_2)\r\,,
\end{split}
\ee
we can compute the auto-covariances of the estimators~\eqref{eq:realest},
\be
\begin{split}
& C_{k_ik_j}=\frac{2}{
\mathcal{N}_{k_i}
}\delta_{ij}P^2(k_i)\,,\quad  C_{TT'}=\frac{(2\pi)^3\pi s_{123}}{k_1k_2k_3\Delta k^3 V}\delta_{TT'}\prod_{i=1}^3 
P(k_i)\,,
\end{split} 
\ee
where $s_{123}=~$6, 2 or 1 for equilateral, isosceles and general triangles.
The cross-covariance $\langle P(k') B(k_1,k_2,k_3) \rangle$ is given by,
\be
 C_{k'_i T}= \frac{2(2\pi)^3}{
 \mathcal{N}_{k'_i}
 }\left(\delta_{ij_1}P(k'_i)B(k_{j_1},k_{j_2},k_{j_3})+\text{cycl.}\right)
\ee
It is straightforward to generalize these calculations to
power spectrum multipole $\ell$ and the redshift-space bispectrum multipole $\ell'$,
\be
\begin{split}
& \hat P_\ell(k_i) = \int_{q\in k_i~\text{shell}} \frac{d^3q}{(2\pi)^3{\cal N}_k }~\delta_0(-\q) \delta_0({\q})(2\ell+1)\mathcal{L}_\ell(\hat{\z}\cdot \hat{\q})
% \,,\quad {\cal N}_k = 4\pi k^2 \Delta k \frac{V}{(2\pi)^3} 
\,, \\
& \hat B_{\ell'}(k_1,k_2,k_3) = (2\ell'+1)\prod_{i=1}^3\int_{k_i} \frac{d^3q_i}{(2\pi)^3}~\frac{(2\pi)^3\delta^{(3)}_D(\q_{123})}{N_T^{123}}
\delta(\q_1)\delta(\q_2)\delta(\q_3)\mathcal{L}_{\ell'}(\hat{\z}\cdot \hat{\q}_1)\,,
\end{split}
\ee
where $\delta_0(\k)=\delta(\k)(1+\beta \mu^2)$ in the linear approximation~\cite{Kaiser:1987qv}, $\beta\equiv f/b_1$.
In particular, this implies that the continuous part of the angle-averaged bispectrum  
auto-covariance would be modulated in redshift space by a form-factor
\be 
\int\frac{d\phi}{2\pi}\int_0^1d\mu~(1+\beta \mu^2)^2(1+\beta \mu_2(\mu,\phi)^2)^2(1+\beta \mu_3(\mu,\phi)^2)^2 \,.
\ee
Similarly, the cross-correlation between $P_\ell$ and $B_0$ is given by
\be 
\begin{split}
& \langle P_\ell(k) B_0(k_1,k_2,k_3) \rangle
\\&=\frac{2(2\pi)^3(2\ell+1)}{\mathcal{N}_{k}} \left(\delta_{k k_1}P(k)\int\frac{d\phi}{2\pi}\int_0^1d\mu~(1+\beta \mu^2)^2\mathcal{L}_\ell(\mu)
B(\k,\k_2,\k_3)+\text{cycl.}\right)\,.
% (1+\beta \mu_2(\mu,\phi)^2)^2(1+\beta \mu_3(\mu,\phi)^2)^2
\end{split}
\ee

\section{Gaussian fingers-of-God exponent derivation}
\label{app:fog}

In this section we revisit the derivation 
of the Gaussian FoG exponent 
that is often used in the literature 
to motivate some phenomenological
models for FoG, see e.g.~\cite{Gil-Marin:2014sta}.
Ref.~\cite{Chudaykin:2020hbf} has explicitly shown 
that this model completely fails to capture the behaviour
seen in high quality dark matter redshift space 
simulations. Nevertheless, it would be 
of some interest to see when the Gaussian 
FoG model breaks down at the mathematical level.
Let us consider the redshift space mapping,
\be
\delta^{(z)}(\k)= \int d^3x~\e^{i\k\x+i\mathcal{H}^{-1}k_zv_z(\x)}(\delta(\x)+1)-(2\pi)^3\delta^{(3)}_D(\k)
\ee
Now we split the velocity field into the long and short wavelength components,
\be
v_i= v^L_i + v^S_i\,,
\ee
where $v^L$ is correlated with the density field on large scales 
and $v^S$ is the short-scale contribution generated by the non-perturbative
effects such as virialization. 
A common assumption is that this 
part is fully uncorrelated with the perturbative long wavelength 
density field. 
Taylor-expanding the exponent over its perturbative part we have
\be 
\delta^{(z)}(\k)= \int d^3x ~\e^{i\k\x+i\mathcal{H}^{-1}k_zv^S_z(\x)}(\delta(\x)+i\mathcal{H}^{-1}k_zv^L_z(\x))\,,
\ee
where we have neglected terms 
which have support only around $\k=0$. 
In what follows we restrict ourselves 
to the tree-level order for the perturbative part, in which case the 
velocity field can be expressed as 
\be 
v^L_i= -f\mathcal{H}\frac{\d_i \delta^L_m}{\Delta}\quad \Rightarrow\quad 
i\mathcal{H}^{-1}k_zv^L_i(\k)=f\mu^2 \delta^L_m(\k)\,.
\ee
In order to reproduce the Gaussian FoG exponent, we need to assume the short scale velocity field is Gaussian 
distributed, and its two point correlation function has a finite support 
on short scales, 
\be 
\begin{split}
\langle v_i(\x)v_j(\x')\rangle = \delta_{ij}\H^2\sigma^2_v \quad \text{for}\quad \x=\x' \quad \text{and}\quad 0\quad \text{otherwise}.
\end{split}
\ee 
Clearly, this assumption cannot be justified 
within the EFT approach, which requires
that the short-scale averages 
should depend on all possible operators involving 
low-energy 
degrees of freedom compatible with IR symmetries 
of large-scale structure~\cite{Baumann:2010tm}.
Nevertheless, if we proceed
using the cumulant expansion theorem
\be
\langle \exp\{i X\} \rangle 
=\exp \left\{\sum_{N=1}^\infty \frac{i^N}{N!} \langle X^N\rangle_c \right\}\,,
\ee
we find the power spectrum in redshift space given by 
\be
\begin{split}
\langle \delta^{(z)}(\k)\delta^{(z)}(\k')\rangle = 
(2\pi)^3\delta_D^{(3)}(\k+\k')~\e^{-\sigma_v^2 \mu^2 k^2}P_{\rm Kaiser}(k)\,,
\end{split} 
\ee
where 
\be 
P_{\rm Kaiser}(k)=(b_1+f\mu^2)^2 P(k)+\frac{1}{\bar n}\,.
\ee
For the redshift space bispectrum we have 
\be
\begin{split}
&\langle \delta^{(z)}(\k_1)\delta^{(z)}(\k_2)\delta^{(z)}(\k_3)\rangle = 
(2\pi)^3\delta_D^{(3)}(\k_{123})~\e^{-\frac{\sigma_v^2}{2} 
\sum_{a=1}^3 \mu_a^2 k_a^2
}B_{\rm tree}(\k_1,\k_2,\k_3)\,,
\end{split} 
\ee
which would formally coincide with the leading order EFT expression used in this work
if we Taylor expand the damping exponent and identify
\be 
c_1=\frac{(k_{\rm NL}^r)^2}{2}\sigma_v^2 b_1\,,\quad c_2=\frac{(k_{\rm NL}^r)^2}{2}\sigma_v^2 f\,.
\ee

\bibliography{/Users/michalychforever/Dropbox/my_notes/Q0/short}
\bibliographystyle{JHEP}

\end{document}